# Investigating the short-term effects of particulate matter (PM) chemical components on mortality and the potential modifying effect of extreme temperature: A time-series analysis in London

Xiaolu Zhang[1], Anna Font[2], Anja Tremper[1], Max Priestman[1], Shawn Y. Lee[1], David C. Green[1,3], Dimitris Evangelopoulos[1]*[‡], Gang I. Chen[1]*[‡]

[1]MRC Centre for Environment and Health, Environmental Research Group, Imperial College London, 86 Wood Lane, London, W12 0BZ, UK

[2]IMT Nord Europe, Europe, Institut Mines-Télécom, Univ. Lille, Centre for Education, Research and Innovation in Energy Environment (CERI EE), 59000 Lille, France

[3]HPRU in Environmental Exposures and Health, Imperial College London, UK

*Correspondence to: Dimitris Evangelopoulos (d.evangelopoulos@imperial.ac.uk) and Gang I. Chen (gang.chen@imperial.ac.uk).

[‡]Joint last authors

# Abstract

Particulate matter (PM) is linked to adverse health outcomes, yet the roles of specific PM components and their modification by extreme temperature remain unclear. We examined short-term associations between ten PM chemical components and daily mortality in Greater London (2015-2018). PM components include inorganic aerosols (black carbon from wood burning (BCwb) and traffic exhaust (BCtr), $SO_4$, $NO_3$, and $NH_4$) and organic aerosols (hydrocarbon-like organic aerosol (HOA), biomass burning OA (BBOA), cooking-like OA (COA), more and less oxidized oxygenated OA (MO-OOA and LO-OOA)). We applied quasi-Poisson generalized additive models and weighted quantile sum (WQS) regression to estimate single-pollutant, multi-pollutant, and mixture effects, respectively, and included interaction terms to test effect modification by heat waves and cold spells. All ten components showed positive associations with all-cause mortality in single-pollutant models with stronger estimated risks for respiratory mortality, particularly for $NH_4$, $NO_3$, $SO_4$. In mixture analyses, the WQS index was significantly associated with all-cause mortality (RR = 1.015, 95% CI: 1.006-1.024 per $25^{th}$-percentile increase) and showed a marginally significance with respiratory mortality (RR = 1.018, 95% CI: 0.994-1.042). MO-OOA and COA contributed most to all-cause mortality, while BBOA and BC Wood dominated respiratory effects. Heat waves consistently amplified respiratory risks in both single-pollutant and mixture models with little evidence for cardiovascular mortality. Overall, MO-OOA demonstrated harmful associations across outcomes, suggesting potential toxicity link to secondary atmospheric oxidation processes. These findings support source-specific control strategies and highlight the importance of accounting for extreme temperature in air pollution mitigation policies.

**Highlights:**

- Ten source-specific PM components positively associated with all-cause mortality with stronger effects on respiratory mortality
- Mixture analysis identified MO-OOA and COA as major contributors to all-cause mortality
- BBOA and BC Wood were key drivers of respiratory mortality risks from PM mixtures
- MO-OOA demonstrated consistent and robust associations across all mortality outcomes and models
- Heat waves amplified respiratory risks in both single-pollutant and mixture models

# 1. Introduction

Ambient air pollution remains a major environmental and public health concern, which causes more than 4 million premature deaths annually according to World Health Organization ([WHO, 2021](#)). Particulate matter (PM) is considered as one of the most harmful components due to its small size and heterogeneous compositions. PM consists of diverse chemical components with different origins as well as vastly distinct toxicities ([Kelly & Fussell, 2012](#)). Typically, PM comprises both inorganic aerosols (IA) and organic aerosols (OA) from different emission sources ([Chen, 2022](#)). IA includes primary particles such as sea salt from ocean spray, mineral dust from construction and road surfaces, and black carbon (BC) from incomplete fossil fuel combustion, residential heating and wildfires. Secondary inorganic aerosols (SIA) such as ammonium nitrate ($NH_4NO_3$) and ammonium sulphate ($NH_4SO_4$) are formed through chemical reactions involving ammonia ($NH_3$) from agricultural activities, sulphur dioxide ($SO_2$) from coal burning and shipping exhaust, and nitrogen oxides ($NO_x$) from traffic and industrial emissions. Organic aerosols have various origins, therefore complex compositions. In general, it can be separate into primary organic aerosols (POA) and secondary organic aerosols (SOA). POA originates from direct emissions such as traffic exhaust (hydrocarbon-like OA (HOA)), cooking (COA), and biomass burning (BBOA). SOA forms when volatile organic compounds (VOC) released by both anthropogenic and biogenic sources undergo atmospheric oxidation and subsequently condense into particle phase ([Chen, 2022](#); [Seinfeld & Pandis, 1998](#)). Depending on the oxidation state, SOA can be generally separated into more-oxidised oxygenated OA (MO-OOA) and less-oxidised oxygenated OA (LO-OOA).

To date, most epidemiological studies on PM have primarily focused on total $PM_{2.5}$ or $PM_{10}$ mass,

with exposure assessed either by various instruments or estimates from air quality models. However, this approach may overlook different health effects from PM components, since the concentration and corresponding toxicity may vary significantly depending on their sources and atmospheric conditions (Daellenbach et al., 2020; Franklin et al., 2008; Piper et al., 2024). The advances in aerosol measurement techniques, such as aerosol mass spectrometry (AMS) (Jayne et al., 2000), or aerosol chemical speciation monitors (ACSM) (Fröhlich et al., 2013; Ng et al., 2011), have enabled the measurements of typical PM chemical components, like $NO_3$, $SO_4$, $NH_4$ and OA.

A systematic review by Rohr and Wyzga (2012), reviewing 51 epidemiological and toxicological studies, concluded that short-term exposure to specific PM components may be more strongly associated with adverse health outcomes than total PM mass. The most frequently studied components mainly focused on the typical sources such as $NO_3$, $SO_4$, carbonaceous species, and metals (Rohr & Wyzga, 2012). For instance, Piper et al. (2024) found statistically significant associations between mortality and elemental carbon (EC) and organic carbon (OC) with 0.51% and 0.45% increase of relative risks (RR) per interquartile range (IQR) increase of EC and OC, respectively (Piper et al., 2024). Similarly, Klemm et al. (2011) identified $NO_3$ as the most significant component associated with total mortality in Atlanta (Klemm et al., 2011), and Peng et al. (2009) reported consistent associations with increased emergency hospitalisations for cardiovascular and respiratory conditions across 119 U.S. counties (Peng et al., 2009).

Investigations targeting specific OA sources remain limited with few long-term data available. OA has been reported to contribute between 20% and 90% of total PM mass in Europe (Chen et al., 2022; Crippa et al., 2014). However, identifying specific OA sources, such as HOA, BBOA, COA, and secondary OA species, typically require expensive AMS-like instrument and source apportionment analyses, which is both labour- and cost-intense, therefore they are not widely available in epidemiological studies. As highlighted by Mauderly & Chow (2008), OA species are often embedded within total PM mass measurements without detailed speciation, as a result, its source-specific health effect is still rarely examined and might be underestimated in the existing research (Mauderly & Chow, 2008). Meanwhile, statistical approaches used to assess effects of source-specific pollutant largely rely on single-pollutant models. Given the context of multiple, highly inter-correlated PM components in the real world, this approach may be limited to identify their independent or joint effects.

Another important factor to consider is temperature. Emerging evidence suggests that extreme events such as heat waves and cold spells may both intensify air pollution levels and associated health risks (Anenberg et al., 2020; Areal et al., 2022). Given the increasing frequency of extreme temperature in the UK and across Europe in recent years due to climate change (Kueh & Lin, 2020), there is an urgent need to investigate whether extreme temperature may affect the health impacts of specific PM components.

To address these gaps, sophisticated and advanced multivariate statistical approaches, specifically positive matrix factorisation (PMF), were used to quantitatively deconvolute OA into distinct sources by following a standardized protocol, resolving time series of BBOA, HOA, and COA, MO-OOA and LO-OOA in London from 2015-2018 (Chen et al., 2022). BC were further categorized into two sources, wood burning (BC Wood) and traffic exhaust (BC Traffic) based on the aethalometer model (Sandradewi et al., 2008; Zotter et al., 2016). Together with other SIAs (NH$_4$, NO$_3$, and SO$_4$) measured by ACSM, it enabled us to resolve PM into ten chemical components in total. Comprehensive epidemiological statistical methods were applied by single-pollutant, multi-pollutant, and weighted quantile sum (WQS) regression to evaluate their associations with all-cause and cause-specific mortality, identifying key contributing emission sources and exploring the potential effect modification by extreme temperature. This study based on data collected between 23 November 2015 and 22 February 2018 in Greater London area.

# 2. Methods

**2.1 Health data**

Mortality data were retrieved from the Office for National Statistics (ONS) and were based on death occurrence. The dataset included daily counts of all-cause mortality, along with stratified information by sex (male and female) and age group (<65 and ≥65 years old). Cause-specific mortality was classified according to the International Classification of Diseases 10$^{th}$ Revision (ICD-10) (WHO, 2024), focusing on cardiovascular deaths (ICD-10 codes: I00-I99) and respiratory deaths (ICD-10 codes: J00-J99).

**2.2 Exposure assessment**

The exposure data for PM components were obtained from the North Kensington air quality monitoring station (51.521050 N, 0.213419 W), part of the London Air Quality Network (LAQN, 2025). This site is classified as an urban background monitoring location, surrounded by residential buildings and positioned away from major roads and industrial areas. It is therefore considered to provide a representative estimate of typical ambient exposure for the general urban population in London. More detailed information of this site can be found: https://uk-air.defra.gov.uk/networks/site-info?site_id=KC1.

The dataset included daily concentrations of ten chemical components of PM, measured in micrograms per cubic metre (μg/m$^3$). These included BC measured using Aethalometer 31 with 1-min time resolution, which was further categorised into wood burning (BC Wood) and traffic sources (BC Traffic) from black carbon source apportionment model (Sandradewi et al., 2008;

Zotter et al., 2016). ACSM was used to measure OA and secondary inorganic aerosols including $NO_3$, $SO_4$, and $NH_4$ in 30-min time resolution. OA were further quantified into five species by rolling PMF, including hydrocarbon-like OA (HOA), biomass burning OA (BBOA), cooking OA (COA), and two secondary oxygenated OA including more oxidised-oxygenated OA (MO-OOA) and less oxidised-oxygenated OA (LO-OOA), originating from both long-range transported airmass and local production of chemical reactions among gas and particulate pollutants (Chen et al., 2022).

All measurements were systematically ratified through strict quality assurance and quality control process by the Aerosol Science Team at Imperial College London and were subsequently aggregated into daily average concentrations. Air pollution measurements were available for most of the study period (November 2015 to February 2018), though some days had missing data due to instrument failures.

**2.3 Meteorological data**

Meteorological data were obtained from the Integrated Surface Database (ISD) maintained by NOAA (NCEI, 2026), using observations from the London Heathrow Airport station. Hourly temperature (°C) and relative humidity (RH; %) data were aggregated to derive daily mean, maximum, and minimum temperature, and daily mean RH. The data is available throughout the overall study period.

**2.4 Different definitions of extreme temperature**

To assess the potential modifying effect of extreme temperatures, five alternative methods were applied to define heat waves and cold spells based on commonly used definitions reported in previous epidemiological studies (Vaidyanathan et al., 2016; Wilson et al., 2013; Xie et al., 2013). All methods were based on percentile thresholds of daily maximum temperature (Tmax) and minimum temperature (Tmin) and required a minimum duration of consecutive days.

- **Method 1** defined heat waves as periods of at least two consecutive days with Tmax exceeding the 95$^{th}$ percentile of the entire study period, and cold spells as periods with Tmin below the 5$^{th}$ percentile.

- **Method 2** used the same percentile thresholds as Method 1 but required a minimum of three consecutive days.

- **Method 3** applied a less extreme cutoff, using the 90$^{th}$ percentile for Tmax and 10$^{th}$ percentile for Tmin, with a minimum of two consecutive days.

- **Method 4** used the same percentile thresholds as Method 3 but required three consecutive days, further limiting the number of identified events.

- **Method 5** used monthly percentile thresholds (90$^{th}$ for Tmax and 10$^{th}$ for Tmin) rather than percentiles calculated from the entire study period, identifying events of at least two consecutive days.

Method 3 was selected for the main analysis due to its consistent absolute temperature across the board, ensuring less variability on PM compositions among extreme events, while Method 5 was used in sensitivity analyses as it reflects seasonal temperature variability and potential adaptation effects from a monthly perspective (**Table S1**).

**2.5 Statistical analysis plan**

Summary statistics table was calculated for all key variables, including daily number of deaths, concentrations of ten PM chemical components, total $PM_{2.5}$ mass and meteorological variables. Time-series plots were generated to visualise seasonal patterns across PM components. A Pearson correlation matrix was used to assess collinearities among pollutants, temperature and humidity. For the five methods used to identify extreme temperature events, we summarised the total number of days classified as heat waves or cold spells, and the corresponding mean PM concentrations and mortality during the study period. Days with missing values in PM components, temperature or relative humidity were excluded from statistical analyses.

Quasi-Poisson generalized additive models (GAMs) were used to examine the associations, as they allow for the modelling of non-linear relationships and account for overdispersion in daily mortality counts (Hastie, 1992; Ver Hoef & Boveng, 2007). We first conducted single-pollutant analyses, with each of the ten PM chemical components examined individually in relation to all-cause, cardiovascular, and respiratory mortality. To evaluate potential attenuation by overall particle mass, multi-pollutant GAM models were then fitted with adjustment for total $PM_{2.5}$ mass. However, given the high correlations among PM components, we further applied weighted quantile sum (WQS) regression to assess their joint effects (Carrico et al., 2015). The WQS model constructs an index that combines all ten PM components into a single mixture, assigning a weight to each component that are aggregated to one. The models were fitted under a positive direction, with the dataset randomly divided into 40% for training and 60% for validation. The component weights were estimated through 10,000 bootstrap samples, and each component was categorised into four quantiles (25% increments) for the main analysis and ten quantiles (10% increments) for sensitivity analyses. To evaluate potential effect modification by extreme temperatures, interaction terms between each PM component and extreme temperature were extended into the single-pollutant models and WQS models. Moreover, distributed lag non-linear models (DLNM) were applied to assess delayed effects (Gasparrini, 2011), examining both single-day lags (lag 0 to lag 6) and the cumulative lag effect over lag 0-6. Stratified analyses were conducted by season (warm: May-October; cold: November-April), age group (<65 vs. ≥65 years old), and sex (male vs. female) to assess potential differences in vulnerability across population subgroups and environmental context.

All fitted models were adjusted for potential confounders, including the non-linear effects of temperature and RH using smooth splines with 5 and 4 degrees of freedom (df) respectively. Long-term and seasonal trends were controlled using a natural spline of time with df of 6 per year. Other confounders included day of the week (categorical) and public holidays (binary). The selection of spline parameters was based on prior literature (Bhaskaran et al., 2013; Piper et al., 2024), exploratory plots and model fit criteria by Akaike Information Criterion (AIC) (Akaike, 1974).

For sensitivity analyses, df for spline terms were varied including temperature (df = 5 to 6), relative humidity (df = 4 to 5), and date (df = 6 to 7 per year). An alternative definition of extreme temperature based on monthly temperature distribution was applied to assess the robustness of effect modification of extreme events (Method 5). Different set of quantiles (q = 10) in WQS were applied to evaluate the stability of component weights and mixture effect estimates.

All statistical analyses were conducted using R version 4.4.1. Relative risks (RR) and 95%

confidence intervals (CI) were used to quantify associations between PM component concentrations and health outcomes. For single- and multi-pollutant models, effect estimates were reported per interquartile range (IQR) increase in PM component concentrations. In the WQS model, effect estimates were interpreted per one quantile increase in the mixture index (i.e., a 25% quantile increase when q = 4; a 10% quantile increase when q = 10). Model diagnostics included assessment of overdispersion using the ratio of residual deviance to degrees of freedom, and evaluation of residual autocorrelation using partial autocorrelation function (PACF) plots.

## 3. Results

### 3.1 Descriptive analysis

Descriptive statistics of air pollutants and mortality outcomes are presented in **Table 1**, covering an overall of 823 total days. Total $PM_{2.5}$ had a mean concentration of (11.48±9.63) µg/m³. Among the PM components, BC Wood had the lowest average of (0.23±0.27) per µg/m³, whereas $NO_3$ exhibited the highest mean concentration and the greatest variability (2.30±3.09 µg/m³). The mean health outcomes were 138±21 for all-cause mortality, 36±7 for cardiovascular mortality, and 19±7 for respiratory mortality. The majority of deaths occurred among individuals aged ≥65 years. For meteorological variables, the daily mean temperature varied between -0.62°C and 25.84°C, with an average value of 11.37°C. Relative humidity ranged from 48.31% to 100%, with a mean of 78.11%.

Table 1. Descriptive summary table of PM components and health outcome data collected for London between 23 November 2015 and 22 February 2018 (n = 823 days).

|  | Availability | Mean | SD | Minimum | Median | Maximum | IQR |
|---|---|---|---|---|---|---|---|
| **Particulate Pollutants** | | | | | | | |
| (24h Avg, µg/m3) | | | | | | | |
| $PM_{2.5}$ | 95% | 11.48 | 9.63 | 2.17 | 8.21 | 102.19 | 7.70 |
| BC Wood | 85% | 0.23 | 0.27 | 0.00 | 0.16 | 2.90 | 0.17 |
| BC Traffic | 85% | 0.71 | 0.68 | 0.05 | 0.53 | 5.84 | 0.48 |
| $NH_4$ | 72% | 1.02 | 1.05 | 0.00 | 0.67 | 8.13 | 0.95 |
| $NO_3$ | 72% | 2.30 | 3.09 | 0.06 | 1.11 | 22.07 | 2.46 |
| $SO_4$ | 72% | 0.72 | 0.60 | 0.02 | 0.58 | 3.91 | 0.73 |
| HOA | 85% | 0.47 | 0.56 | 0.04 | 0.30 | 4.53 | 0.31 |
| BBOA | 85% | 0.56 | 0.58 | 0.04 | 0.40 | 5.32 | 0.47 |
| COA | 85% | 0.59 | 0.46 | 0.07 | 0.46 | 3.76 | 0.46 |
| MO-OOA | 85% | 1.20 | 1.22 | 0.13 | 0.81 | 11.90 | 0.98 |
| LO-OOA | 85% | 0.95 | 1.28 | 0.07 | 0.59 | 18.28 | 0.83 |
| **Mortality Data** | | | | | | | |
| All-Cause Mortality | 100% | 138 | 21 | 93 | 135 | 202 | 29 |
| Female | 100% | 70 | 12 | 38 | 69 | 105 | 16 |
| Male | 100% | 68 | 12 | 38 | 67 | 108 | 16 |
| Under 65 | 100% | 27 | 6 | 14 | 27 | 81 | 8 |
| Over and equal 65 | 100% | 111 | 19 | 71 | 108 | 168 | 25 |
| Cardiovascular Mortality | 100% | 36 | 7 | 16 | 36 | 63 | 10 |
| Respiratory Mortality | 100% | 19 | 7 | 4 | 17 | 48 | 8 |
| **Meteorological Data** | | | | | | | |
| (24h Avg) | | | | | | | |
| Temperature (°C) | 100% | 11.37 | 5.48 | -0.62 | 11.13 | 25.84 | 8.69 |
| Relative Humidity (%) | 100% | 78.11 | 10.01 | 48.31 | 78.87 | 100.00 | 14.57 |

Time-series plots (**Figure S1**) demonstrated clear seasonal variation in PM components, with pronounced winter peaks observed in early 2016, late 2016, and early 2017. Consistent with these patterns, seasonal stratification showed higher concentrations during the cold season (November-April) compared with the warm season (May-October) for most components (**Table S2**). When further classified by extreme temperature events, pollutant concentrations were generally higher during cold spells than during heat waves, except for $SO_4$, which consistently showed higher levels during heat waves across all temperature definitions (**Table S3**).

Pearson correlation (**Table 2**) examined relationships among all PM components, total $PM_{2.5}$ mass, and meteorological data. Most components showed strong correlations with total $PM_{2.5}$, with $NH_4$ (r = 0.86), $NO_3$ (r = 0.86), and LO-OOA (r = 0.83) exhibiting the highest values. The strongest correlations were observed between $NH_4$ and $NO_3$ (r = 0.96), BC Traffic and HOA (r = 0.92), and BC Wood and BBOA (r = 0.87). Temperature was negatively correlated with most pollutants, and correlations between pollutants and relative humidity were generally weaker.

1   Table 2. Pearson correlation coefficients among PM chemical components, total PM$_{2.5}$ mass and meteorological data for London between 23 November 2015 and 22 February 2018.
2   Bolded values indicate relatively strong correlations (r ≥ 0.85).

|  | BC Wood | BC Traffic | NH$_4$ | NO$_3$ | SO$_4$ | HOA | BBOA | COA | MO-OOA | LO-OOA | PM$_{2.5}$ | Temperature | RH |
|---|---|---|---|---|---|---|---|---|---|---|---|---|---|
| **BC Wood** | 1.00 | - | - | - | - | - | - | - | - | - | - | - | - |
| **BC Traffic** | 0.84 | 1.00 | - | - | - | - | - | - | - | - | - | - | - |
| **NH$_4$** | 0.51 | 0.56 | 1.00 | - | - | - | - | - | - | - | - | - | - |
| **NO$_3$** | 0.53 | 0.56 | **0.96** | 1.00 | - | - | - | - | - | - | - | - | - |
| **SO$_4$** | 0.19 | 0.31 | 0.70 | 0.60 | 1.00 | - | - | - | - | - | - | - | - |
| **HOA** | **0.85** | **0.92** | 0.56 | 0.56 | 0.31 | 1.00 | - | - | - | - | - | - | - |
| **BBOA** | **0.87** | 0.81 | 0.61 | 0.60 | 0.31 | **0.85** | 1.00 | - | - | - | - | - | - |
| **COA** | 0.66 | 0.71 | 0.56 | 0.54 | 0.39 | 0.76 | 0.78 | 1.00 | - | - | - | - | - |
| **MO-OOA** | 0.45 | 0.50 | 0.83 | 0.84 | 0.68 | 0.52 | 0.57 | 0.56 | 1.00 | - | - | - | - |
| **LO-OOA** | 0.79 | 0.77 | 0.67 | 0.63 | 0.52 | 0.81 | 0.77 | 0.65 | 0.61 | 1.00 | - | - | - |
| **PM$_{2.5}$** | 0.74 | 0.77 | **0.86** | **0.86** | 0.62 | 0.74 | 0.73 | 0.65 | 0.80 | 0.83 | 1.00 | - | - |
| **Temperature** | -0.42 | -0.27 | -0.24 | -0.35 | 0.23 | -0.25 | -0.31 | -0.09 | -0.09 | -0.12 | -0.28 | 1.00 | - |
| **RH** | 0.30 | 0.30 | 0.20 | 0.24 | -0.08 | 0.18 | 0.23 | 0.09 | 0.04 | 0.10 | 0.22 | -0.49 | 1.00 |

**3.2 Single-pollutant regression analysis**

**Figure 1** presents the estimated RR with 95% CI for the associations between individual PM component and mortality outcomes based on single-pollutant models. For all-cause mortality, positive associations were observed for all components (**Figure 1a**). The strongest significant associations were found for $SO_4$ (RR = 1.017, 95% CI: 1.006-1.029), MO-OOA (RR = 1.014, 95% CI: 1.008-1.020), $NH_4$ (RR = 1.012, 95% CI: 1.004-1.020), NO3 (RR = 1.011, 95% CI: 1.004-1.018), and COA (RR = 1.012, 95% CI: 1.004-1.020). The remaining components also showed positive effects, with lower confidence limits marginally close to one. No statistically significant associations were observed between PM components and cardiovascular mortality (**Figure 1b)**.

For respiratory mortality, effect estimates were generally larger but accompanied by wider confidence intervals (**Figure 1c**). $SO_4$ had the largest effect estimate (RR = 1.025, 95% CI: 0.997-1.054), followed in magnitude by $NH_4$ (RR = 1.023, 95% CI: 1.004-1.042) and $NO_3$ (RR = 1.017, 95% CI: 1.001-1.035). Several other components, including BBOA, COA, and MO-OOA also demonstrated approximately 1.5% increases in risk per IQR with marginally significant associations. In comparison, BC Wood and BC Traffic showed relatively weaker effects.

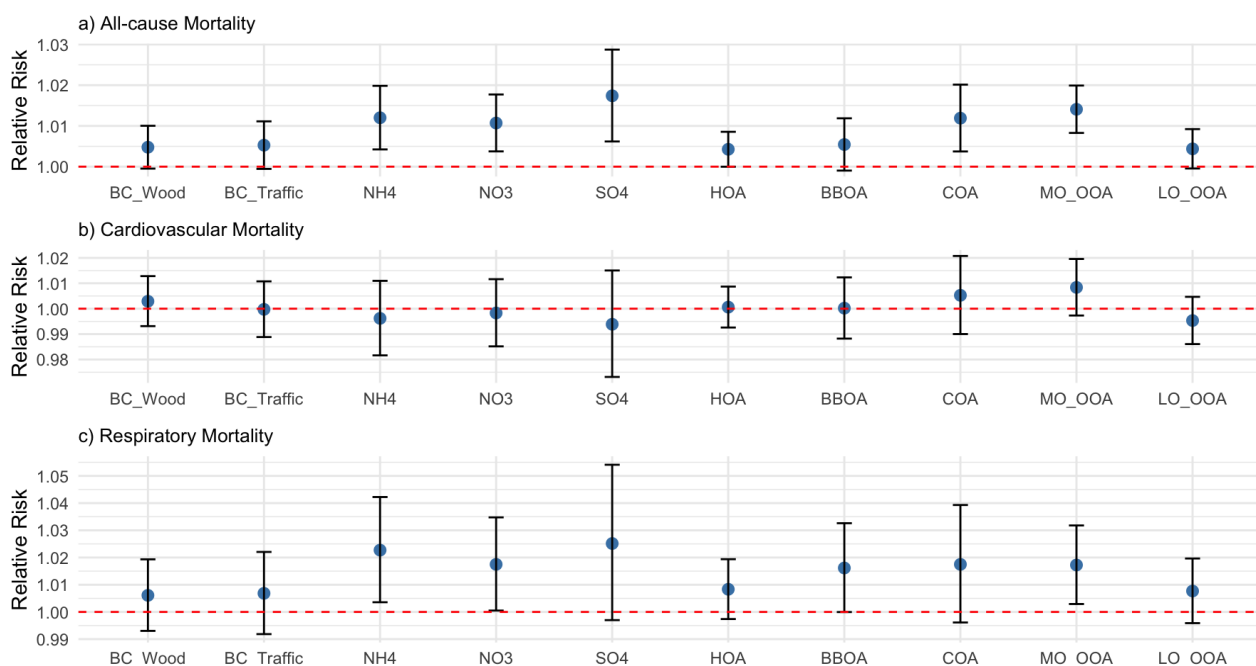

**Figure 1.** Relative risks (RR) and 95% confidence intervals (CI) for the association between interquartile range (IQR) increases in individual PM components and mortality outcomes. Results are shown for: (a) all-cause mortality, (b) cardiovascular mortality, and (c) respiratory mortality. Models were adjusted for temperature, RH, long-term and seasonal trends, weekdays, and public holidays. The red dashed line indicates RR = 1.00.

### 3.3 Multi-pollutant regression analysis

After adjusting for total $PM_{2.5}$ mass, effect estimates for most PM components were attenuated and their confidence intervals became wider across all mortality outcomes (**Figure 2**). For all-cause mortality, nearly all components showed reduced associations after adjustment, with the attenuation particularly evident for BC Wood, BC Traffic and LO-OOA. A similar pattern was observed for respiratory mortality where the effect estimates were reduced for several components except for $NH_4$ and MO-OOA with slight increase in risk estimates (ΔRR = 0.62% and 1.05%, respectively). Notably, for cardiovascular mortality, although most components remained non-significant, MO-OOA demonstrated the most pronounced risk and became statistically significant after adjustment (ΔRR = 1.12%, Δ95% CI: 0.31%- 1.96%).

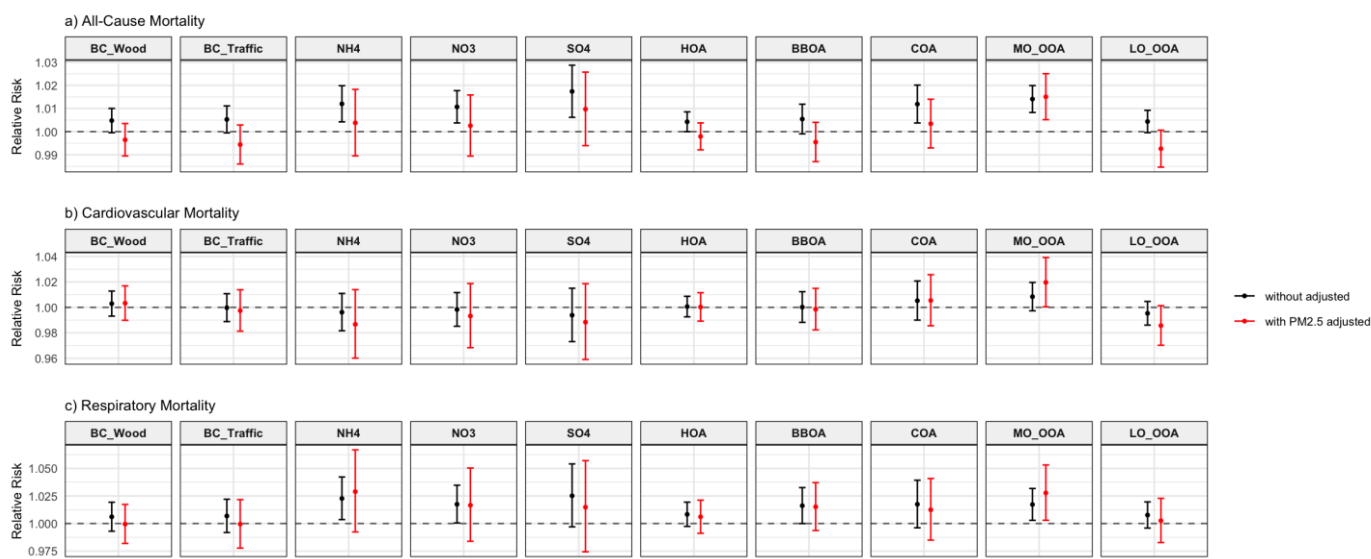

**Figure 2.** Relative risks (RR) and 95% confidence intervals (CI) for the associations between interquartile range (IQR) increases in individual PM components and mortality outcomes, estimated from single-pollutant models without adjustment (black) and with additional adjustment for total PM$_{2.5}$ mass (red). Results are shown for: (a) all-cause mortality, (b) cardiovascular mortality, and (c) respiratory mortality. All models were adjusted for temperature, relative humidity (RH), long-term and seasonal trends, weekdays, and public holidays. The red dashed line indicates RR = 1.00.

### 3.4 WQS regression analysis

Considering the high correlation between PM components, WQS regression was applied to evaluate the joint effects of ten components as a mixture (**Table 3**). The overall mixture index was positively associated with all-cause mortality, corresponding to an RR of 1.015 (95% CI: 1.006-1.024) per 25 percentiles increase in the WQS index. The largest weights were assigned to MO-OOA (25%), COA (20%), BC Traffic (14%), and BC Wood (14%). A relatively weak association was observed for cardiovascular mortality for the mixture index (RR = 1.003, 95% CI: 0.985-1.021). In terms of respiratory mortality, the mixture effect was larger in magnitude and borderline significant (RR = 1.018, 95% CI: 0.994-1.042). As indicated by their mixture weights, BBOA and BC Wood contributed the largest proportions to respiratory outcome, accounting for 27% and 17% of the total weight respectively.

When specifying q = 10 in the WQS model, the effect estimates were proportional to the 25th quantile increment by approximately 2.5 times smaller as expected, remaining positive and marginally significant for all-cause and respiratory mortality (**Table S4**). Component weights varied to some extent, but the dominant contributors remained similar.

Table 3. WQS regression results for the association between the mixture of ten PM components and mortality outcomes. Mixture effect estimates are expressed as relative risks (RR) with 95% confidence intervals (CI) per 25th percentile increase in the WQS index (q = 4). Component weights represent the mean contribution of each pollutant to the mixture index, estimated using 10,000 bootstrap samples. The WQS models were fitted using 40% of the data for training and 60% for validation.

| All-cause Mortality | | Cardiovascular Mortality | | Respiratory Mortality | |
|---|---|---|---|---|---|
| **Mixture effect** | 1.015 | **Mixture effect** | 1.003 | **Mixture effect** | 1.018 |
| *(RR with 95%CI)* | (1.006, 1.024) | *(RR with 95%CI)* | (0.985, 1.021) | *(RR with 95%CI)* | (0.994, 1.042) |
| **Weight** *(% with 95% CI)* | | **Weight** *(% with 95% CI)* | | **Weight** *(% with 95% CI)* | |
| MO-OOA | 25% (3%, 64%) | BC Wood | 31% (4%, 72%) | BBOA | 27% (2%, 72%) |
| COA | 20% (1%, 62%) | COA | 23% (1%, 59%) | BC Wood | 17% (1%, 52%) |
| BC Traffic | 14% (1%, 38%) | MO-OOA | 20% (0%, 59%) | $SO_4$ | 14% (0%, 59%) |
| BC Wood | 14% (0%, 49%) | LO-OOA | 6% (0%, 27%) | $NH_4$ | 12% (0%, 36%) |
| LO-OOA | 8% (0%, 28%) | HOA | 5% (0%, 28%) | MO-OOA | 9% (0%, 36%) |
| $SO_4$ | 8% (0%, 35%) | BC Traffic | 5% (0%, 27%) | COA | 7% (0%, 31%) |
| BBOA | 4% (0%, 23%) | BBOA | 5% (0%, 18%) | BC Traffic | 4% (0%, 18%) |
| $NH_4$ | 4% (0%, 15%) | $SO_4$ | 4% (0%, 17%) | LO-OOA | 4% (0%, 19%) |
| $NO_3$ | 3% (0%, 14%) | $NH_4$ | 2% (0%, 12%) | HOA | 3% (0%, 17%) |
| HOA | 2% (0%, 12%) | $NO_3$ | 1% (0%, 7%) | $NO_3$ | 2% (0%, 11%) |

**3.5 Effects of extreme temperatures**

Effect modification analyses based on the single-pollutant models are presented in **Figure S2**. No clear pattern of modification was observed for all-cause mortality. Risk estimates during cold spells were similar to the overall results, and several components such as BBOA, MO-OOA, LO-OOA exhibited slightly elevated estimates under heat waves, but with wider confidence intervals. For cardiovascular mortality, associations remained weak with confidence intervals spanning unity under extreme temperature events in both single-pollutant and WQS models.

In contrast, the modifying effect was particularly pronounced for respiratory mortality during heat waves, but with larger variation. All PM components were associated with increased risks under heat waves with only BC Wood falling narrowly statistical significance. The most significant associations were observed for BBOA (RR = 1.209, 95% CI: 1.095-1.335), $NH_4$ (RR = 1.184, 95% CI: 1.071-1.309), $NO_3$ (RR = 1.183, 95% CI: 1.061-1.321) and $SO_4$ (RR = 1.155, 95% CI: 1.054-1.267).

Modification patterns were further validated by including an interaction term in the WQS regression (**Table S5**). For all-cause mortality, the mixture effect was strengthened during cold spells (RR = 1.030, 95% CI: 0.989-1.073), whereas little change was observed during heat waves (RR = 1.001, 95% CI: 0.955-1.049). For cardiovascular mortality, no statistically significant modification was observed, with slightly elevated estimates during cold spells and lower estimates during heat waves, both accompanied by confidence intervals crossing one. Consistent significant patterns were found for respiratory mortality under heat waves, corresponding to a 3.8% increase in risk for every 25 percentiles increase in the WQS index (RR = 1.038, 95% CI: 0.994-1.084).

Effect modification analyses using monthly temperature distributions (Method 5) were broadly consistent with the main analysis (Method 3), with stronger associations observed under heat waves for respiratory mortality across most components such as $SO_4$, $NH_4$ and MO-OOA (**Figure S3**).

**3.6 Stratified analysis**

Seasonal stratification revealed stronger associations for respiratory mortality during the warm season (May-October). Several components, including $NH_4$, $SO_4$, $NO_3$, MO-OOA, and BBOA, demonstrated larger effect estimates (generally >1.05) compared with the cold season (**Figure S4**). No clear evidence of heterogeneity by sex or age was found in the standard GAM model as confidence intervals were largely overlapped (**Figure S5**).

**3.7 Lagged effect analysis**

Lagged associations between PM components and mortality showed significant effects on the day of exposure and in the following 1-2 days, with relative risks generally decreasing after lag 3. Significant cumulative effects over lag 0-6 were also observed, including positive associations for all components with all-cause mortality, and for $NH_4$, $NO_3$, $SO_4$ and MO-OOA with respiratory mortality, and specially for MO-OOA with cardiovascular mortality (**Figure S6-S8**).

When stratifying lagged effects by subgroup, no consistent pattern was evident for single lag days given the substantial overlap in confidence intervals between males and females. Although females appeared to have higher cumulative risks over lag 0-6, estimates were accompanied by wide confidence intervals (**Figure S9**). Similarly, effect estimates for individuals under 65 years tended to be higher but with large uncertainty. The strongest cumulative association was observed for COA (RR = 1.040, 95% CI: 1.007-1.073) (**Figure S10**).

For the sensitivity analysis on varying the degrees of freedom for temperature (df = 5-6), relative humidity (df = 4-5), and long-term trends (df = 6-7 per year), the associations between PM components and mortality outcomes remained stable (**Tables S6-S8**). Model diagnostics indicated no substantial overdispersion and minimal residual autocorrelation, with residual patterns supporting adequate model fit (**Table S9; Figures S11-S12**).

# 4. Discussion

**4.1 Interpretation of main findings**

Our study provides comprehensive evidence on the short-term health effects of ten PM chemical components in London between November 2015 and February 2018. Nearly all components were significantly associated with all-cause mortality from single-pollutant models, and the associations were generally stronger in magnitude for respiratory mortality, particularly for secondary inorganic aerosols such as $NH_4$, $NO_3$, and $SO_4$ with no significant association found for cardiovascular mortality.

The adverse health effects of $NO_3$ and $SO_4$ in our study were aligned with several previous studies across multiple health outcomes (Hang et al., 2022; Klemm et al., 2011; Weichenthal et al., 2021). In our study, $NO_3$ accounted for the largest proportion of total $PM_{2.5}$ mass (20%), mainly originating from the oxidation of $NO_x$ emitted by traffic exhaust, while $SO_4$ primarily originates from long-range transported airmasses and ship exhaust emissions. Laboratory experiment has demonstrated that exposure to both particulate $NO_3$ and $SO_4$ can impair lung function and promote inflammatory responses such as neutrophil infiltration (Zhang et al., 2021). However, despite epidemiological evidence indicating their health risks, toxicological studies have yet to confirm the independent harmful effects of $NO_3$ and $SO_4$ under ambient conditions (Reiss et al., 2007; Sagai, 2019).

Evidence regarding the health effects of $NH_4$ in prior studies is relatively limited to best of authors' knowledge. $NH_4$ are formed through atmospheric reactions between ammonia ($NH_3$) and acidic precursors such as $NO_x$ and $SO_2$ (Seinfeld & Pandis, 1998). In the UK and across Europe, $NH_3$ emissions primarily stem from agricultural sources such as livestock and fertilizer use (AQEP, 2018; EEA, 2017). Also, northern France and the Benelux are the hot spots of $NH_4$ with more intensive agricultural activities (Schneider et al., 2025; Sebnem et al., 2020) and often chemically stable, therefore, it is often associated with long-range transport airmasses coming from continental Europe (especially in Spring). These sustained agricultural emission sources may therefore contribute to elevated ambient ammonium levels and potential adverse health outcomes during the study period. Considering $NH_4$ and $NO_3$ are highly correlated (r = 0.96 in our dataset), In addition, their effects may reflect shared regional sources rather than independent toxicities. This supports the need for mixture-based methods to better characterise the health impacts of correlated pollutants.

After adjustment for total $PM_{2.5}$ mass, most component-specific associations were attenuated and lost statistical significance as expected, likely due to the strong correlations between individual components and overall particle mass. Notably, MO-OOA remained robust across outcomes in the muti-pollutant models and even became significant for cardiovascular mortality, whereas LO-OOA showed smaller and less consistent associations. Although both are SOA formed through atmospheric photochemical processes, MO-OOA is often more oxidised and stayed in the atmosphere for longer time. Meanwhile, MO-OOA exhibited stronger correlations with regional secondary pollutants such as $NH_4$ (r = 0.84) and $NO_3$ (r = 0.85), suggesting a shared influence from long-range transported and atmospherically aged air masses (Lhotka et al., 2025; Via et al., 2021; Vieno et al., 2014). Experimental evidence further indicates that highly oxidized organic aerosols such as MO-OOA may induce reactive oxygen species formation, supporting a potential oxidative stress pathway (Liu et al., 2023). In contrast, LO-OOA is more influenced by local emissions (Chen et al., 2022; Lhotka et al., 2025). In our descriptive analysis, LO-OOA showed stronger correlations with POA sources (e.g., HOA and COA), consistent with the study Lhotka et al. (2025) that LO-OOA is largely influenced by local origin and short-range transport (Lhotka et al., 2025). MO-OOA, unlike LO-OOA, typically exhibits greater solubility and smaller particle size as a result of prolonged atmospheric aging and higher oxidation levels (Liangou et al., 2022; Pang et al., 2006; Qiu et al., 2019). These physicochemical properties may contribute to the stronger health effects linked to MO-OOA in this study.

Interestingly, when considering all components as a mixture in a real-world context, COA emerged as a major contributor to all-cause mortality, accounting for 20% of the total mixture weight immediately after the top MO-OOA, whereas $NO_3$, $SO_4$, and $NH_4$ showed relatively less contributions compared with their strong associations in single-pollutant models. One potential explanation is that single-pollutant models may overestimate the effect since the high correlation would capture the variation to increase or decrease from other components, however, the WQS approach assigns weights based on each component's contribution to the overall mixture effect and allows for the assessment of individual contributions under a highly correlated exposure.

Differences in emission sources may also play a role on the health effect. Given that COA emission are primarily emitted from both residential and commercial cooking activities, this is a notable contributor to urban PM particularly in densely populated areas (Reyes-Villegas et al., 2018; G. Zhang et al., 2023). However, in London, most of COA is primarily generated from commercial kitchens (Chen et al., 2025). Although epidemiological evidence on COA remains limited compared to other components, our findings on its contribution to all-cause mortality can add to a growing recognition on its public health relevance. COA is typically composed of primary organic particles that are rich in unsaturated fatty acids and other reactive compounds, which may undergo atmospheric aging and contribute to secondary aerosol formation (Mohr et al., 2009; Takhar et al., 2021). Given its prevalence in urban environments (6% of PM) and potential oxidative properties (Chen et al., 2022), further research is needed to better understand its toxicological mechanisms.

BBOA and BC Wood accounted for the largest contributions to the respiratory mixture effect (27% and 17%, respectively). Their strong correlation (r = 0.87) suggests a shared origin from wood burning predominately for residential heating purposes, which is a major source in urban environments. Experimental and panel studies have shown that short-term exposure to wood smoke can induce airway inflammation and reduced lung function in humans (Barregard et al., 2008; Riddervold et al., 2012). Given that BBOA contains a large fraction of water-soluble organic carbon (WSOC), which has been linked to increased oxidative potential in ambient particulate matter (Cao et al., 2021; Duarte & Duarte, 2021). From a mechanistic perspective, water-soluble components can absorb moisture once inhaled into the humid respiratory tract, causing particles to grow in size (Varghese & Gangamma, 2006). This change in particle size influences where particles deposit within the lungs, which could help explain stronger associations with respiratory outcomes.

In sensitivity analyses using q = 10 in the WQS model, the dominant contributors remained while the relative ranking of component contributions shifted. For example, the top contributors under q = 4 were MO-OOA (25%), COA (20%), BC Traffic (14%), and BC Wood (14%) for all-cause mortality, whereas under q = 10, BC Traffic (24%) and BC Wood (21%) ranked highest, followed by MO-OOA (19%) and COA (10%). Such variation in ranking may reflect differences in quantile categorisation, as each PM component does not increase linearly with PM increases, which leads to nonlinearity of contribution to RR across different quantile levels. However, the key components remained among the leading contributors, supporting the robustness of the mixture findings.

**4.2 Interpretation of effect modification**

Cold spells did not exhibit a clear modification pattern in the single-pollutant models; however, the mixture analysis indicated a modest increase in all-cause mortality, with a marginally significant association (RR = 1.030, 95% CI: 0.989-1.073). Notably, most PM components exhibited higher concentrations during cold spells, particularly for $NO_3$, which was dominant in the overall mass contribution and is known to remain more stable in particulate form under low temperature (Stelson & Seinfeld, 1982). Increased exposure under such conditions, combined with heightened physiological vulnerability during cold weather, may together contribute to the amplified health effects observed during cold spells (Zeb et al., 2024; Q. Zhang et al., 2023).

Heat waves significantly amplified the association with respiratory mortality, with consistent evidence from both single-pollutant and WQS models. Seasonal stratification further revealed stronger respiratory associations during the warm season compared with the effects during both cold season and all days. From a biological perspective, heat exposure is known to induce dehydration, vasodilation and increased pulmonary ventilation, all of which may weaken respiratory function and enhance pollutant inhaled (Andersen et al., 2023; Epstein & Yanovich, 2019). Among the effect modification on all components under heat waves, BBOA exhibited the strongest respiratory health effect under heat wave conditions with the highest RR of 1.209 with every IQR increase of BBOA. However, these findings should be considered with caution, since the contributions of BBOA tend to be relatively low during heat waves. These may potentially increase measurement uncertainty and variability in source apportionment estimates. In addition, BBOA showed high correlations (r>0.79) with $PM_{2.5}$ and other major PM components (i.e., BC Traffic, HOA, MO-OOA, and LO-OOA). Thus, it might be the mixture effect instead of the risk from BBOA alone.

The limited associations observed with cardiovascular mortality in this study is possibly because cardiovascular outcomes are more strongly influenced by long-term cumulative exposure rather than short-term effects (Beelen et al., 2014; Brook et al., 2010). This limits the ability of time-series studies to detect such associations. Previous research has suggested that the cardiovascular effects of PM are primarily driven by chronic low-dose exposure, which can lead to systemic inflammation, endothelial dysfunction and other long-term pathophysiological changes (Beelen et al., 2014).

**4.3 Interpretation of stratified analysis**

In stratified analyses by sex, we observed a relative higher cumulative effect among females. From a biological perspective, women tend to have smaller airway diameters and lower lung volumes compared to men, which may lead to higher pollutant deposition per unit body weight (LoMauro & Aliverti, 2018; Sturm, 2016). However, current evidence on sex differences in the health effects of air pollution remains inconsistent (Hwashin Hyun et al., 2022), and our estimates were accompanied with wide and overlapping confidence intervals. Therefore, the observed associations should be interpreted with caution, and further research is required to explore the potential sex differences in susceptibility.

For age-stratified analyses, individuals under 65 years exhibited relatively higher cumulative mortality risks over lag 0-6, possibly because that younger populations may engage in more frequent outdoor activities (e.g., commuting, exercise, social activities), which can increase personal exposure to ambient air pollution and thus cumulative inhaled dose ([Gao et al., 2025](); [Wu et al., 2024]()).

**4.4 Public health implications**

Source-specific interventions could play a critical role. For example, international joint efforts to reduce $NH_4$ pollution should target ammonia emissions from agriculture, mainly through better fertilizer and manure management, especially since much of it originates from continental Europe. Controlling $NO_3$ pollution requires the promotion of low-emission transportation options, as well as cleaner and electrical cook stoves instead of gas stoves. Given that COA originates from local commercial cooking, the government needs to establish stricter regulations to reduce emissions using filtration system instead of relying on dispersion, such as UV-C lamp hoods to break down oil molecules, or electrostatic precipitators to capture oil particles ([HKEPD, 2024]()). Similarly, reducing BBOA and BC Wood concentrations calls for stricter regulation of residential wood burning and other biomass combustion sources, including improved stove standards, the promotion of cleaner heating alternatives, and limited solid fuel use for commercial cooking activities ([Chen et al., 2025]()).

As global temperatures continue to rise and heat waves become more frequent in recent years, susceptible populations, particularly individuals with pre-existing respiratory conditions such as asthma, should be advised to limit outdoor exposure during extreme heat events to reduce the risk of symptom exacerbation. At the same time, raising public awareness about specific pollution sources (e.g., commercial cooking, wood burning, traffic emissions) and promoting preventive guidelines, such as the use of air purifiers and wearing protective masks, when necessary, may help reduce induvial exposure during extreme events.

**4.5 Strengths and Limitations**

One of the key strengths of this study is the detailed exposure assessment. By employing a high-resolution source apportionment approach, we were able to capture nearly all major PM chemical components, including detailed subcomponents of organic aerosols. This allowed for a more comprehensive and precise quantification of source-specific exposure on health effects.

Another strength lies in the comprehensive statistical approaches, including single-pollutant models, multi-pollutant adjustment, and mixture analysis using WQS regression. The single-pollutant models allowed us to characterise component-specific associations, while adjustment for total $PM_{2.5}$ mass helped evaluate the independence of individual effects. The WQS approach further enabled the assessment of joint mixture effects and identified the main contributors under high collinearity.

In terms of the limitations, our exposure data came from a single monitoring site in London, which may not fully reflect variations in pollutant concentrations across the city. While the North Kensington station provides a good estimate of daily exposures for typical resident in London, it may not capture the neighboured variations, especially for pollutants like BC that vary more across space (Bell et al., 2007; Yongde et al., 2024). In addition, the relatively short study period (from late 2015 to early 2018) may limit the statistical power to capture a longer-term exposure trend. Future research should incorporate data from multiple monitoring sites to better capture spatial variability in exposure and extending the study period to help capturing longer-term exposure patterns.

# 5. Conclusion

In conclusion, our findings provide evidence that short-term effects of several PM chemical components in London are associated with all-cause and respiratory mortality. Considering both individual component effects and broader mixture effects, MO-OOA demonstrated consistent associations across all health outcomes. The mixture analysis further identified COA as a major contributor to all-cause mortality, while residential heating and wood-burning sources (BBOA and BC Wood) were more strongly linked to respiratory mortality. Under extreme temperature events, heat waves may further amplify respiratory risks, highlighting the need for source-specific interventions and greater attention to the interaction between air pollution and extreme temperature. Eventually, this study provides first-ever evidences in short-time health impacts of specific PM sources and their interactions with extreme temperatures, which provide valuable information for policymakers to design more effective abatement policies in a warming climate.

# Supplementary Information for:

# Investigating the short-term effects of particulate matter (PM) chemical components on mortality and the potential modifying effect of extreme temperature: A time-series analysis in London

Xiaolu Zhang[1], Anna Font[2], Anja Tremper[1], Max Priestman[1], Shawn Y. Lee[1], David C. Green[1,3], Dimitris Evangelopoulos[1]*[‡], Gang I. Chen[1]*[‡]

[1]MRC Centre for Environment and Health, Environmental Research Group, Imperial College London, 86 Wood Lane, London, W12 0BZ, UK

[2]IMT Nord Europe, Europe, Institut Mines-Télécom, Univ. Lille, Centre for Education, Research and Innovation in Energy Environment (CERI EE), 59000 Lille, France

[3]HPRU in Environmental Exposures and Health, Imperial College London, UK

*Correspondence to: Dimitris Evangelopoulos (d.evangelopoulos@imperial.ac.uk) and Gang I. Chen (gang.chen@imperial.ac.uk).

[‡]Joint last authors

# List of Tables

# List of Figures

Table S1. Definitions of extreme temperature events identified using five alternative methods. Each method applied different percentile thresholds and consecutive duration. The table shows the number of events, total days classified, and the percentage of study days meeting each definition for both heat waves and cold spells.

| Method | Definitions | Type | Events (n) | Total days (n) | Percentage (%) |
|---|---|---|---|---|---|
| 1 | Tmax > 95$^{th}$ percentile of entire study period for ≥2 consecutive days | Heat wave | 10 | 35 | 4% |
| 1 | Tmin < 5$^{th}$ percentile of entire study period for ≥2 consecutive days | Cold Spell | 8 | 29 | 4% |
| 2 | Tmax > 95$^{th}$ percentile of entire study period for ≥3 consecutive days | Heat wave | 7 | 29 | 4% |
| 2 | Tmin < 5$^{th}$ percentile of entire study period for ≥3 consecutive days | Cold Spell | 4 | 21 | 3% |
| 3 | Tmax > 90$^{th}$ percentile of entire study period for ≥2 consecutive days | Heat wave | 17 | 70 | 9% |
| 3 | Tmin < 10$^{th}$ percentile of entire study period for ≥2 consecutive days | Cold Spell | 15 | 61 | 7% |
| 4 | Tmax > 90$^{th}$ percentile of entire study period for ≥3 consecutive days | Heat wave | 13 | 62 | 8% |
| 4 | Tmin < 10$^{th}$ percentile of entire study period for ≥3 consecutive days | Cold Spell | 11 | 53 | 6% |
| 5 | Tmax > 90$^{th}$ percentile of the monthly distribution for ≥2 consecutive days | Heat wave | 21 | 52 | 6% |
| 5 | Tmin < 10$^{th}$ percentile of the monthly distribution for ≥2 consecutive days | Cold Spell | 15 | 36 | 4% |

Table S2. Summary statistics of mortality counts and PM component concentrations during warm season (May to October) and cold season (November to April). Values are presented as mean (standard deviation). Bolded values indicate the higher value between warm season and cold season.

| Type | All-Cause Mortality (n) | Cardiovascular Mortality (n) | Respiratory Mortality (n) |
|---|---|---|---|
| Warm season | 124 (14) | 33 (6) | 14 (4) |
| Cold season | **149 (19)** | **39 (7)** | **22 (7)** |

| Type | Total $PM_{2.5}$ (µg/m³) | BC Wood (µg/m³) | BC Traffic (µg/m³) | $NH_4$ (µg/m³) | $NO_3$ (µg/m³) | $SO_4$ (µg/m³) | HOA (µg/m³) | BBOA (µg/m³) | COA (µg/m³) | MO-OOA (µg/m³) | LO-OOA (µg/m³) |
|---|---|---|---|---|---|---|---|---|---|---|---|
| Warm season | 9.69 (6.12) | 0.15 (0.11) | 0.64 (0.50) | 0.91 (0.69) | 1.73 (1.97) | **0.91 (0.50)** | 0.41 (0.40) | 0.45 (0.35) | **0.62 (0.44)** | 1.13 (0.89) | 0.93 (0.85) |
| Cold season | **13.00 (11.62)** | **0.29 (0.33)** | **0.77 (0.79)** | **1.33 (1.25)** | **3.51 (3.64)** | 0.73 (0.67) | **0.51 (0.66)** | **0.65 (0.70)** | 0.57 (0.47) | **1.26 (1.44)** | **0.97 (1.53)** |

Table S3. Summary statistics of mortality counts and PM component concentrations during heat waves and cold spell days identified by five different temperature definitions (Methods 1-5). Values are presented as mean (standard deviation). Bolded values indicate the higher value between heat wave and cold spell days under the same method.

| Method | Type | All-Cause Mortality (n) | Cardiovascular Mortality (n) | Respiratory Mortality (n) |
|---|---|---|---|---|
| 1 | Heat wave | 130 (16) | 33 (6) | 15 (5) |
| 1 | Cold Spell | **165 (18)** | **44 (7)** | **27 (8)** |
| 2 | Heat wave | 133 (14) | 34 (6) | 16 (5) |
| 2 | Cold Spell | **163 (19)** | **43 (8)** | **27 (8)** |
| 3 | Heat wave | 129 (18) | 33 (7) | 15 (5) |
| 3 | Cold Spell | **160 (17)** | **43 (7)** | **26 (6)** |
| 4 | Heat wave | 128 (18) | 34 (6) | 15 (5) |
| 4 | Cold Spell | **159 (17)** | **42 (7)** | **25 (6)** |
| 5 | Heat wave | **146 (19)** | **38 (7)** | **20 (8)** |
| 5 | Cold Spell | 137 (21) | 36 (8) | 18 (6) |

| Method | Type | Total PM$_1$ (µg/m³) | BC Wood (µg/m³) | BC Traffic (µg/m³) | NH$_4$ (µg/m³) | NO$_3$ (µg/m³) | SO$_4$ (µg/m³) | HOA (µg/m³) | BBOA (µg/m³) | COA (µg/m³) | MO-OOA (µg/m³) | LO-OOA (µg/m³) |
|---|---|---|---|---|---|---|---|---|---|---|---|---|
| 1 | Heat wave | 11.22(5.90) | 0.15 (0.08) | 0.81 (0.46) | 1.12 (0.72) | 1.8 (1.71) | **1.34 (0.64)** | 0.56 (0.32) | 0.63 (0.33) | 0.98 (0.51) | 2.02 (1.18) | 1.8 (0.97) |
| 1 | Cold Spell | **29.49(17.54)** | **1.18 (0.72)** | **2.61 (1.70)** | **2.86 (1.99)** | **8.18 (5.75)** | 1.11 (0.85) | **2.11 (1.39)** | **2.15 (1.38)** | **1.62 (0.88)** | **3.23 (1.72)** | **4.45 (4.52)** |
| 2 | Heat wave | 12.53(5.73) | 0.16 (0.08) | 0.92 (0.43) | 1.23 (0.75) | 2.08 (1.77) | **1.4 (0.68)** | 0.63 (0.31) | 0.7 (0.32) | 1.09 (0.49) | 2.3 (1.13) | 2.01 (0.95) |
| 2 | Cold Spell | **32.34(17.93)** | **1.24 (0.75)** | **2.83 (1.76)** | **3.16 (2.07)** | **9.15 (5.89)** | 1.27 (0.86) | **2.26 (1.44)** | **2.18 (1.43)** | **1.71 (0.88)** | **3.56 (1.71)** | **4.99 (4.79)** |
| 3 | Heat wave | 9.64(5.31) | 0.15 (0.08) | 0.75 (0.43) | 0.99 (0.61) | 1.63 (1.52) | **1.2 (0.54)** | 0.57 (0.45) | 0.58 (0.32) | 0.87 (0.51) | 1.66 (1.01) | 1.55 (0.93) |
| 3 | Cold Spell | **20.02(16.95)** | **0.77 (0.67)** | **1.73 (1.53)** | **1.87 (1.75)** | **5.41 (5.08)** | 0.86 (0.71) | **1.33 (1.25)** | **1.44 (1.22)** | **1.11 (0.83)** | **2.11 (1.71)** | **2.64 (3.58)** |
| 4 | Heat wave | 9.53(5.41) | 0.14 (0.08) | 0.73 (0.44) | 0.97 (0.61) | 1.55 (1.47) | **1.21 (0.57)** | 0.51 (0.33) | 0.55 (0.31) | 0.83 (0.46) | 1.64 (1.04) | 1.48 (0.89) |
| 4 | Cold Spell | **19.88(17.66)** | **0.78 (0.70)** | **1.74 (1.61)** | **1.86 (1.85)** | **5.32 (5.36)** | 0.85 (0.75) | **1.36 (1.30)** | **1.5 (1.28)** | **1.13 (0.86)** | **2.16 (1.75)** | **2.73 (3.8)** |
| 5 | Heat wave | 8.83(6.97) | 0.18 (0.12) | 0.69 (0.40) | 1.21 (0.94) | 2.64 (2.85) | **0.99 (0.66)** | 0.46 (0.39) | 0.56 (0.45) | 0.67 (0.49) | **1.57 (1.31)** | 1.18 (1.06) |
| 5 | Cold Spell | **13.55(16.20)** | **0.45 (0.66)** | **1.12 (1.46)** | **1.4 (1.63)** | **3.54 (4.74)** | 0.89 (0.79) | **0.83 (1.19)** | **0.86 (1.20)** | **0.75 (0.79)** | 1.47 (1.59) | **1.82 (3.22)** |

Table S4. WQS regression results for the association between the mixture of ten PM components and mortality outcomes. Mixture effect estimates are expressed as relative risks (RR) with 95% confidence intervals (CI) per 25th percentile increase in the WQS index (q = 10). Component weights represent the mean contribution of each pollutant to the mixture index, estimated using 10,000 bootstrap samples. The WQS models were fitted using 40% of the data for training and 60% for validation.

| **All-cause Mortality** | | **Cardiovascular Mortality** | | **Respiratory Mortality** | |
|---|---|---|---|---|---|
| **Mixture effect** | 1.006 | **Mixture effect** | 1.001 | **Mixture effect** | 1.007 |
| *(RR with 95%CI)* | (1.003, 1.010) | *(RR with 95%CI)* | (0.994, 1.008) | *(RR with 95%CI)* | (0.998, 1.016) |
| **Weight** *(% with 95% CI)* | | **Weight** *(% with 95% CI)* | | **Weight** *(% with 95% CI)* | |
| BC Traffic | 24% (2%, 55%) | BC Wood | 35% (4%, 80%) | BBOA | 31% (2%, 75%) |
| BC Wood | 21% (1%, 67%) | COA | 18% (0%, 52%) | $SO_4$ | 17% (0%, 68%) |
| MO-OOA | 19% (1%, 53%) | MO-OOA | 14% (0%, 54%) | BC Wood | 15% (0%, 46%) |
| COA | 10% (0%, 37%) | BC Traffic | 10% (0%, 46%) | MO-OOA | 9% (0%, 25%) |
| SO4 | 10% (0%, 39%) | HOA | 6% (0%, 32%) | BC Traffic | 9% (0%, 36%) |
| LO-OOA | 7% (0%, 25%) | BBOA | 6% (0%, 30%) | $NH_4$ | 6% (0%, 27%) |
| BBOA | 3% (0%, 15%) | $SO_4$ | 5% (0%, 23%) | LO-OOA | 4% (0%, 21%) |
| $NO_3$ | 2% (0%, 9%) | LO-OOA | 4% (0%, 22%) | COA | 4% (0%, 24%) |
| $NH_4$ | 2% (0%, 12%) | $NH_4$ | 2% (0%, 11%) | $NO_3$ | 3% (0%, 14%) |
| HOA | 1% (0%, 8%) | $NO_3$ | 1% (0%, 8%) | HOA | 2% (0%, 13%) |

**Table S5.** WQS regression results for the association between the mixture of ten PM components and mortality outcomes under extreme temperature events. Mixture effect estimates are presented as relative risks (RR) with 95% confidence intervals (CI) per 25th percentile increase in the WQS index (q = 4). The WQS models included interaction terms between the mixture index and extreme temperatures (heat waves and cold spells). Interaction effects were quantified using the formula: $RR = exp(\beta_{mixture} + \beta_{mixture \times temp})$, and corresponding standard errors were calculated by $SE = \sqrt{SE(\beta_{mixture})^2 + SE(\beta_{mixture \times temp})^2}$. Models were trained using 40% of the dataset and validated on the remaining 60%.

| All-cause Mortality | | Cardiovascular Mortality | | Respiratory Mortality | |
|---|---|---|---|---|---|
| **All Days** | 1.015 (1.006, 1.024) | **All Days** | 1.003 (0.985, 1.021) | **All Days** | 1.007 (0.998, 1.016) |
| **Heat Waves** | 1.001 (0.955, 1.049) | **Heat Waves** | 0.928 (0.848, 1.016) | **Heat Waves** | 1.038 (0.994, 1.084) |
| **Cold Spells** | 1.030 (0.989, 1.073) | **Cold Spells** | 1.058 (0.959, 1.167) | **Cold Spells** | 1.023 (0.984, 1.062) |

**Table S6.** Sensitivity analysis results for **all-cause mortality**. Relative risks (RRs) and 95% confidence intervals (CIs) per interquartile range (IQR) increase in PM component concentrations under different model specifications, changing df for temperature (df = 5-6), relative humidity (df = 4-5), and date (df = 6-7 per year).

| Temperature (df) | Humidity (df) | Date (df per year) | BC Wood | BC Traffic | $NH_4$ | $NO_3$ | $SO_4$ |
|---|---|---|---|---|---|---|---|
| 5 | 4 | 6 | 1.005 (1.000, 1.010) | 1.005 (0.999, 1.011) | 1.012 (1.004, 1.020) | 1.011 (1.004, 1.018) | 1.017 (1.006, 1.029) |
| 5 | 4 | 7 | 1.005 (1.000, 1.010) | 1.005 (0.999, 1.011) | 1.012 (1.004, 1.020) | 1.011 (1.003, 1.018) | 1.016 (1.005, 1.028) |
| 5 | 5 | 6 | 1.005 (0.999, 1.010) | 1.005 (1.000, 1.011) | 1.012 (1.004, 1.020) | 1.011 (1.004, 1.018) | 1.017 (1.006, 1.029) |
| 5 | 5 | 7 | 1.005 (1.000, 1.010) | 1.005 (0.999, 1.011) | 1.012 (1.004, 1.020) | 1.011 (1.003, 1.018) | 1.016 (1.005, 1.028) |
| 6 | 4 | 6 | 1.005 (0.999, 1.010) | 1.005 (0.999, 1.011) | 1.012 (1.004, 1.020) | 1.011 (1.004, 1.018) | 1.017 (1.006, 1.029) |
| 6 | 4 | 7 | 1.005 (1.000, 1.010) | 1.005 (0.999, 1.011) | 1.012 (1.004, 1.020) | 1.011 (1.003, 1.018) | 1.016 (1.005, 1.028) |
| 6 | 5 | 6 | 1.005 (0.999, 1.010) | 1.005 (1.000, 1.011) | 1.012 (1.004, 1.020) | 1.011 (1.004, 1.018) | 1.017 (1.006, 1.029) |
| 6 | 5 | 7 | 1.005 (1.000, 1.010) | 1.005 (0.999, 1.011) | 1.012 (1.004, 1.020) | 1.011 (1.003, 1.018) | 1.016 (1.005, 1.028) |

| Temperature (df) | Humidity (df) | Date (df per year) | HOA | BBOA | COA | MO-OOA | LO-OOA |
|---|---|---|---|---|---|---|---|
| 5 | 4 | 6 | 1.004 (1.000, 1.009) | 1.005 (0.999, 1.012) | 1.012 (1.004, 1.020) | 1.014 (1.008, 1.020) | 1.004 (1.000, 1.009) |
| 5 | 4 | 7 | 1.004 (1.000, 1.008) | 1.005 (0.999, 1.012) | 1.012 (1.003, 1.020) | 1.014 (1.009, 1.020) | 1.004 (0.999, 1.009) |
| 5 | 5 | 6 | 1.004 (1.000, 1.009) | 1.005 (0.999, 1.012) | 1.012 (1.003, 1.020) | 1.014 (1.008, 1.020) | 1.004 (1.000, 1.009) |
| 5 | 5 | 7 | 1.004 (1.000, 1.009) | 1.005 (0.999, 1.012) | 1.011 (1.003, 1.020) | 1.014 (1.008, 1.020) | 1.004 (0.999, 1.009) |
| 6 | 4 | 6 | 1.004 (1.000, 1.009) | 1.005 (0.999, 1.012) | 1.012 (1.004, 1.020) | 1.014 (1.008, 1.020) | 1.004 (1.000, 1.009) |
| 6 | 4 | 7 | 1.004 (1.000, 1.008) | 1.006 (0.999, 1.012) | 1.012 (1.004, 1.020) | 1.014 (1.009, 1.020) | 1.004 (0.999, 1.009) |
| 6 | 5 | 6 | 1.004 (1.000, 1.009) | 1.005 (0.999, 1.012) | 1.012 (1.004, 1.020) | 1.014 (1.008, 1.020) | 1.004 (1.000, 1.009) |
| 6 | 5 | 7 | 1.004 (1.000, 1.009) | 1.005 (0.999, 1.012) | 1.011 (1.003, 1.020) | 1.014 (1.009, 1.020) | 1.004 (0.999, 1.009) |

Table S7. Sensitivity analysis results for **cardiovascular mortality**. Relative risks (RRs) and 95% confidence intervals (CIs) per interquartile range (IQR) increase in PM component concentrations under different model specifications, changing df for temperature (df = 5-6), relative humidity (df = 4-5), and date (df = 6-7 per year).

| Temperature (df) | Humidity (df) | Date (df per year) | BC Wood | BC Traffic | $NH_4$ | $NO_3$ | $SO_4$ |
|---|---|---|---|---|---|---|---|
| 5 | 4 | 6 | 1.003 (0.993, 1.013) | 1.000 (0.989, 1.011) | 0.996 (0.982, 1.011) | 0.998 (0.985, 1.012) | 0.994 (0.973, 1.015) |
| 5 | 4 | 7 | 1.002 (0.992, 1.012) | 0.999 (0.988, 1.010) | 0.995 (0.980, 1.010) | 0.997 (0.984, 1.011) | 0.993 (0.972, 1.014) |
| 5 | 5 | 6 | 1.003 (0.993, 1.013) | 1.000 (0.989, 1.011) | 0.996 (0.982, 1.011) | 0.998 (0.985, 1.012) | 0.994 (0.973, 1.015) |
| 5 | 5 | 7 | 1.002 (0.992, 1.012) | 0.999 (0.988, 1.010) | 0.995 (0.980, 1.010) | 0.997 (0.984, 1.011) | 0.993 (0.972, 1.014) |
| 6 | 4 | 6 | 1.003 (0.993, 1.013) | 1.000 (0.989, 1.011) | 0.996 (0.982, 1.011) | 0.998 (0.985, 1.012) | 0.994 (0.973, 1.015) |
| 6 | 4 | 7 | 1.002 (0.993, 1.012) | 0.999 (0.988, 1.010) | 0.995 (0.980, 1.010) | 0.997 (0.984, 1.011) | 0.993 (0.972, 1.014) |
| 6 | 5 | 6 | 1.003 (0.993, 1.013) | 1.000 (0.989, 1.011) | 0.996 (0.982, 1.011) | 0.998 (0.985, 1.012) | 0.994 (0.973, 1.015) |
| 6 | 5 | 7 | 1.002 (0.993, 1.012) | 0.999 (0.988, 1.010) | 0.995 (0.980, 1.010) | 0.997 (0.984, 1.011) | 0.993 (0.972, 1.014) |

| Temperature (df) | Humidity (df) | Date (df per year) | HOA | BBOA | COA | MO-OOA | LO-OOA |
|---|---|---|---|---|---|---|---|
| 5 | 4 | 6 | 1.001 (0.993, 1.009) | 1.000 (0.988, 1.012) | 1.005 (0.990, 1.021) | 1.008 (0.997, 1.020) | 0.995 (0.986, 1.005) |
| 5 | 4 | 7 | 1.000 (0.992, 1.008) | 0.999 (0.987, 1.012) | 1.004 (0.989, 1.020) | 1.008 (0.997, 1.019) | 0.995 (0.985, 1.004) |
| 5 | 5 | 6 | 1.001 (0.993, 1.009) | 1.000 (0.988, 1.012) | 1.005 (0.990, 1.021) | 1.008 (0.997, 1.020) | 0.995 (0.986, 1.005) |
| 5 | 5 | 7 | 1.000 (0.992, 1.008) | 0.999 (0.987, 1.012) | 1.004 (0.989, 1.020) | 1.008 (0.997, 1.019) | 0.995 (0.985, 1.004) |
| 6 | 4 | 6 | 1.001 (0.993, 1.009) | 1.000 (0.988, 1.012) | 1.005 (0.990, 1.021) | 1.009 (0.997, 1.020) | 0.995 (0.986, 1.004) |
| 6 | 4 | 7 | 1.000 (0.992, 1.008) | 1.000 (0.988, 1.012) | 1.005 (0.989, 1.020) | 1.008 (0.997, 1.020) | 0.995 (0.985, 1.004) |
| 6 | 5 | 6 | 1.001 (0.993, 1.009) | 1.000 (0.988, 1.012) | 1.005 (0.990, 1.021) | 1.009 (0.997, 1.020) | 0.995 (0.986, 1.004) |
| 6 | 5 | 7 | 1.000 (0.992, 1.008) | 1.000 (0.988, 1.012) | 1.005 (0.989, 1.020) | 1.008 (0.997, 1.020) | 0.995 (0.985, 1.004) |

Table S8. Sensitivity analysis results for **respiratory mortality**. Relative risks (RRs) and 95% confidence intervals (CIs) per interquartile range (IQR) increase in PM component concentrations under different model specifications, changing df for temperature (df = 5-6), relative humidity (df = 4-5), and date (df = 6-7 per year).

| Temperature (df) | Humidity (df) | Date (df per year) | BC Wood | BC Traffic | NH$_4$ | NO$_3$ | SO$_4$ |
|---|---|---|---|---|---|---|---|
| 5 | 4 | 6 | 1.006 (0.993, 1.019) | 1.007 (0.992, 1.022) | 1.023 (1.004, 1.042) | 1.017 (1.001, 1.035) | 1.025 (0.997, 1.054) |
| 5 | 4 | 7 | 1.007 (0.994, 1.020) | 1.006 (0.991, 1.021) | 1.022 (1.003, 1.042) | 1.017 (1.000, 1.035) | 1.024 (0.996, 1.053) |
| 5 | 5 | 6 | 1.006 (0.993, 1.019) | 1.006 (0.991, 1.022) | 1.023 (1.004, 1.042) | 1.017 (1.001, 1.035) | 1.025 (0.997, 1.054) |
| 5 | 5 | 7 | 1.006 (0.993, 1.020) | 1.006 (0.991, 1.021) | 1.022 (1.003, 1.042) | 1.017 (1.000, 1.035) | 1.024 (0.996, 1.053) |
| 6 | 4 | 6 | 1.004 (0.991, 1.018) | 1.005 (0.990, 1.020) | 1.021 (1.001, 1.040) | 1.016 (0.998, 1.033) | 1.022 (0.993, 1.051) |
| 6 | 4 | 7 | 1.005 (0.992, 1.019) | 1.005 (0.989, 1.020) | 1.021 (1.001, 1.041) | 1.016 (0.998, 1.034) | 1.021 (0.993, 1.051) |
| 6 | 5 | 6 | 1.004 (0.991, 1.017) | 1.005 (0.990, 1.020) | 1.021 (1.001, 1.040) | 1.016 (0.998, 1.033) | 1.022 (0.993, 1.051) |
| 6 | 5 | 7 | 1.005 (0.991, 1.018) | 1.004 (0.989, 1.020) | 1.021 (1.001, 1.041) | 1.016 (0.998, 1.034) | 1.021 (0.993, 1.051) |

| Temperature (df) | Humidity (df) | Date (df per year) | HOA | BBOA | COA | MO-OOA | LO-OOA |
|---|---|---|---|---|---|---|---|
| 5 | 4 | 6 | 1.008 (0.997, 1.019) | 1.016 (1.000, 1.033) | 1.017 (0.996, 1.039) | 1.017 (1.003, 1.032) | 1.008 (0.996, 1.020) |
| 5 | 4 | 7 | 1.008 (0.997, 1.019) | 1.017 (1.000, 1.033) | 1.017 (0.996, 1.039) | 1.018 (1.004, 1.033) | 1.008 (0.996, 1.020) |
| 5 | 5 | 6 | 1.008 (0.997, 1.019) | 1.016 (1.000, 1.032) | 1.017 (0.996, 1.039) | 1.017 (1.003, 1.032) | 1.007 (0.996, 1.019) |
| 5 | 5 | 7 | 1.008 (0.997, 1.019) | 1.016 (1.000, 1.033) | 1.017 (0.995, 1.038) | 1.018 (1.004, 1.032) | 1.007 (0.995, 1.019) |
| 6 | 4 | 6 | 1.007 (0.996, 1.018) | 1.015 (0.998, 1.031) | 1.015 (0.993, 1.037) | 1.016 (1.002, 1.031) | 1.006 (0.994, 1.018) |
| 6 | 4 | 7 | 1.007 (0.996, 1.018) | 1.015 (0.999, 1.032) | 1.015 (0.994, 1.037) | 1.017 (1.003, 1.032) | 1.006 (0.995, 1.019) |
| 6 | 5 | 6 | 1.007 (0.996, 1.018) | 1.014 (0.998, 1.031) | 1.015 (0.993, 1.037) | 1.016 (1.002, 1.031) | 1.006 (0.994, 1.018) |
| 6 | 5 | 7 | 1.007 (0.995, 1.018) | 1.015 (0.999, 1.032) | 1.015 (0.993, 1.037) | 1.017 (1.003, 1.032) | 1.006 (0.994, 1.018) |

Table S9. Dispersion parameters were calculated for each pollutant-specific model using the Pearson statistic. Values close to 1 indicate no evidence of overdispersion.

| Pollutant | All-Cause Mortality | Cardiovascular Mortality | Respiratory Mortality |
|---|---|---|---|
| BC Wood | 1.16 | 1.11 | 1.05 |
| BC Traffic | 1.17 | 1.11 | 1.05 |
| $NH_4$ | 1.19 | 1.12 | 1.03 |
| $NO_3$ | 1.19 | 1.12 | 1.04 |
| $SO_4$ | 1.19 | 1.12 | 1.04 |
| HOA | 1.16 | 1.11 | 1.04 |
| BBOA | 1.17 | 1.11 | 1.04 |
| COA | 1.16 | 1.11 | 1.04 |
| MO-OOA | 1.13 | 1.10 | 1.04 |
| LO-OOA | 1.17 | 1.10 | 1.04 |

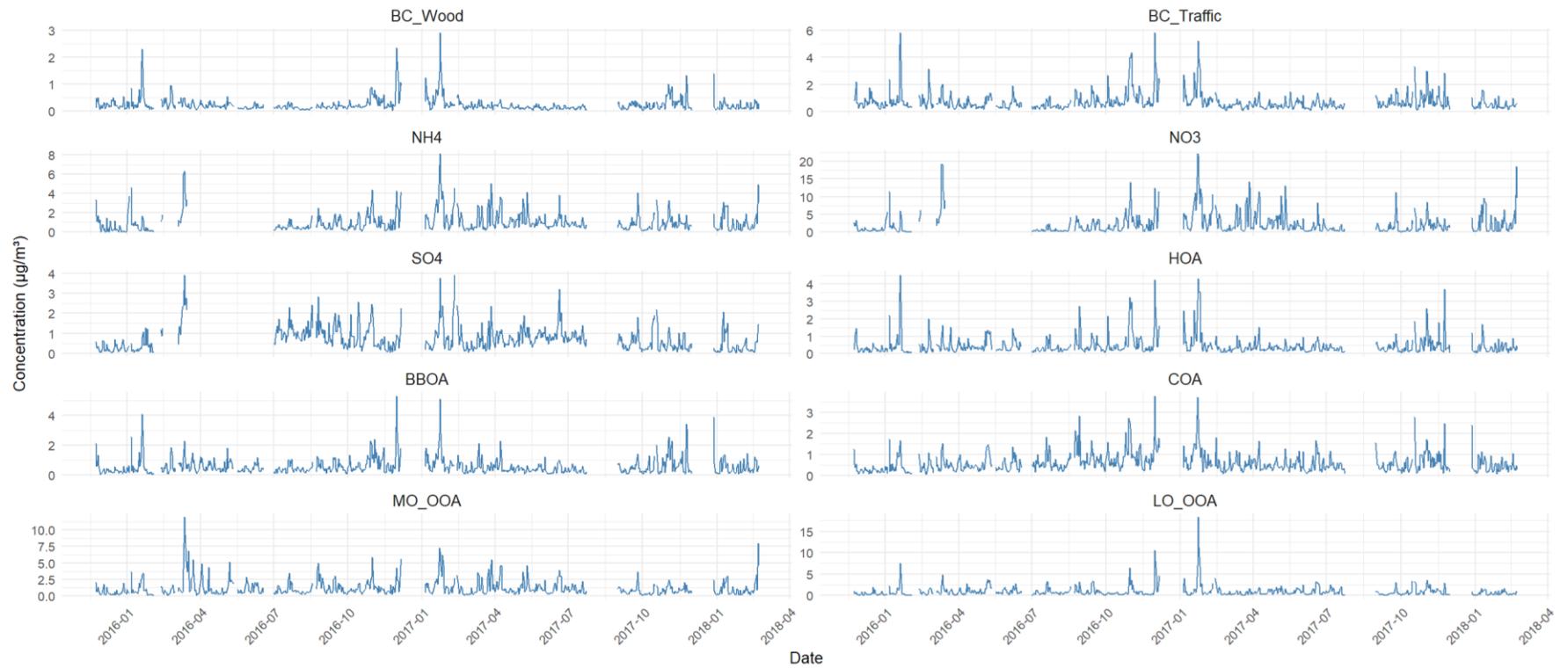

**Figure S1.** Time-series plots for the concentrations of ten PM chemical components in Greater London between 23 November 2015 and 22 February 2018.

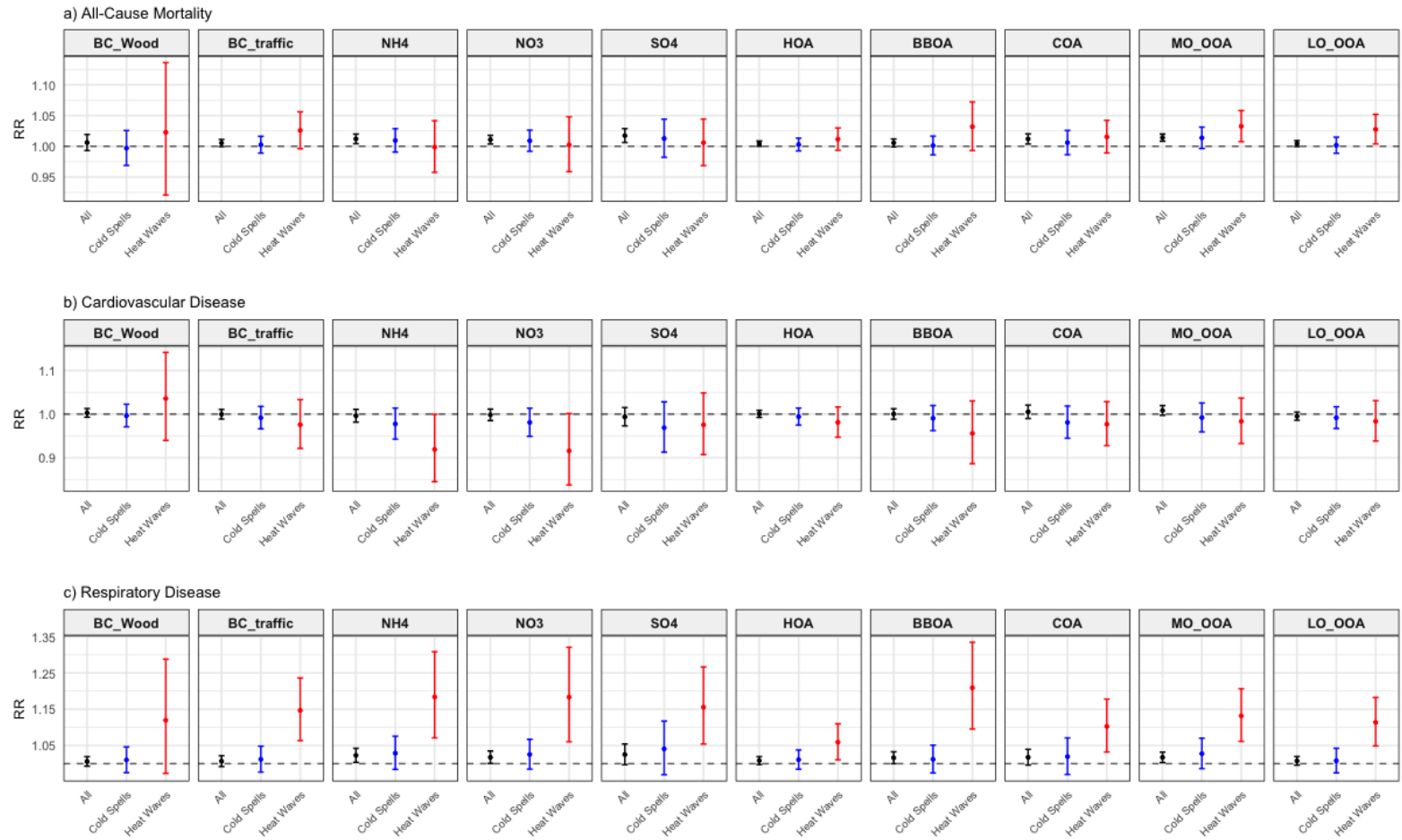

**Figure S2.** Relative risks (RR) and 95% confidence intervals (CI) for the associations between interquartile range (IQR) increases in PM chemical components and daily mortality outcomes (all-cause, cardiovascular, and respiratory), based on interaction models incorporating extreme temperature conditions. Cold spells and heat waves were defined according to Method 3, where heat waves correspond to periods with maximum temperature (Tmax) exceeding the 90th percentile and cold spells to periods with minimum temperature (Tmin) falling below the 10th percentile, each sustained for at least two consecutive days over the entire study period. Models were adjusted for temperature, relative humidity, long-term and seasonal trends, day of the week, and public holidays. The dashed line indicates RR = 1.00.

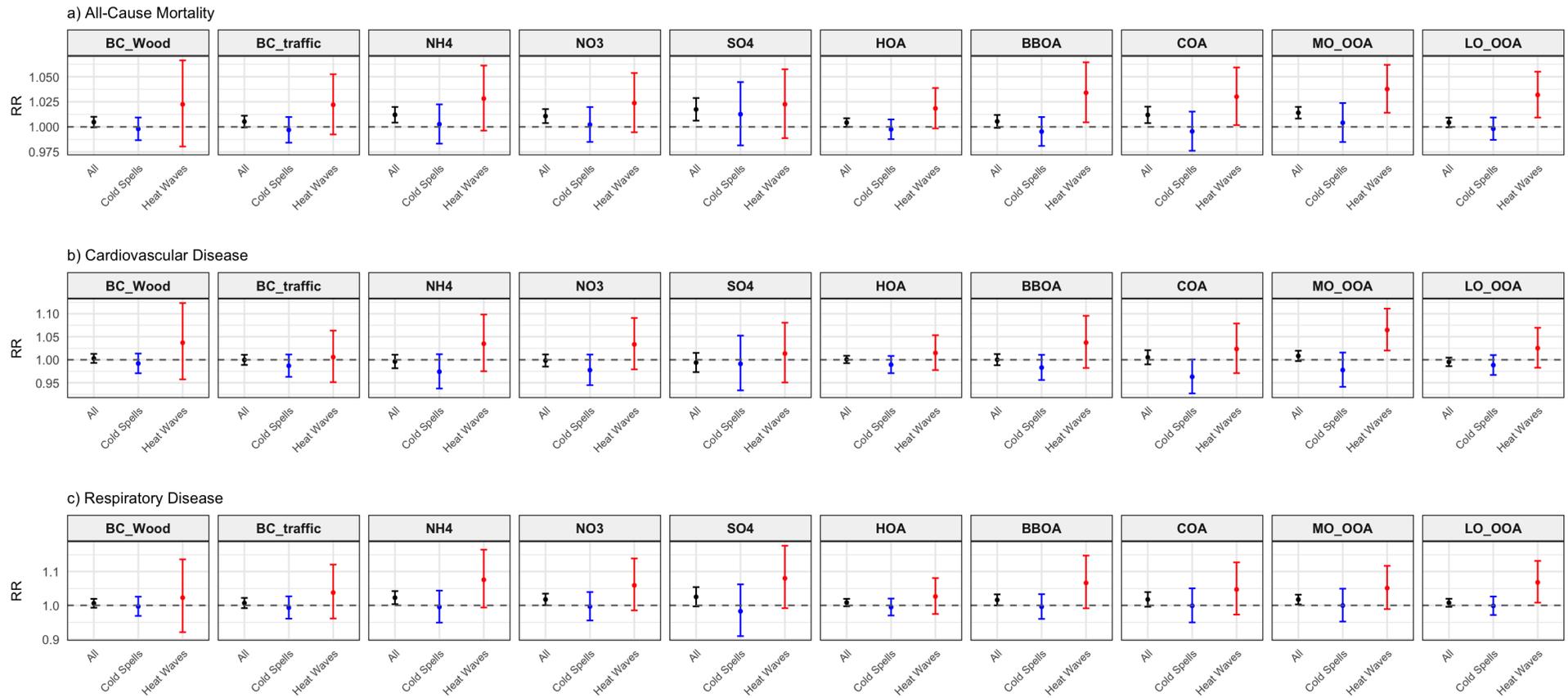

**Figure S3.** Relative risks (RRs) and 95% confidence intervals for the associations between interquartile range (IQR) increases in PM chemical components and daily mortality outcomes (all-cause, cardiovascular, and respiratory), based on interaction models incorporating extreme temperature conditions. Cold spells and heat waves were defined according to Method 5, where heat waves correspond to periods with maximum temperature (Tmax) exceeding the 90th percentile and cold spells to periods with minimum temperature (Tmin) falling below the 10th percentile of the monthly temperature distribution, each sustained for at least two consecutive days. Models were adjusted for temperature, relative humidity, long-term and seasonal trends, day of the week, and public holidays. The dashed line indicates RR = 1.00.

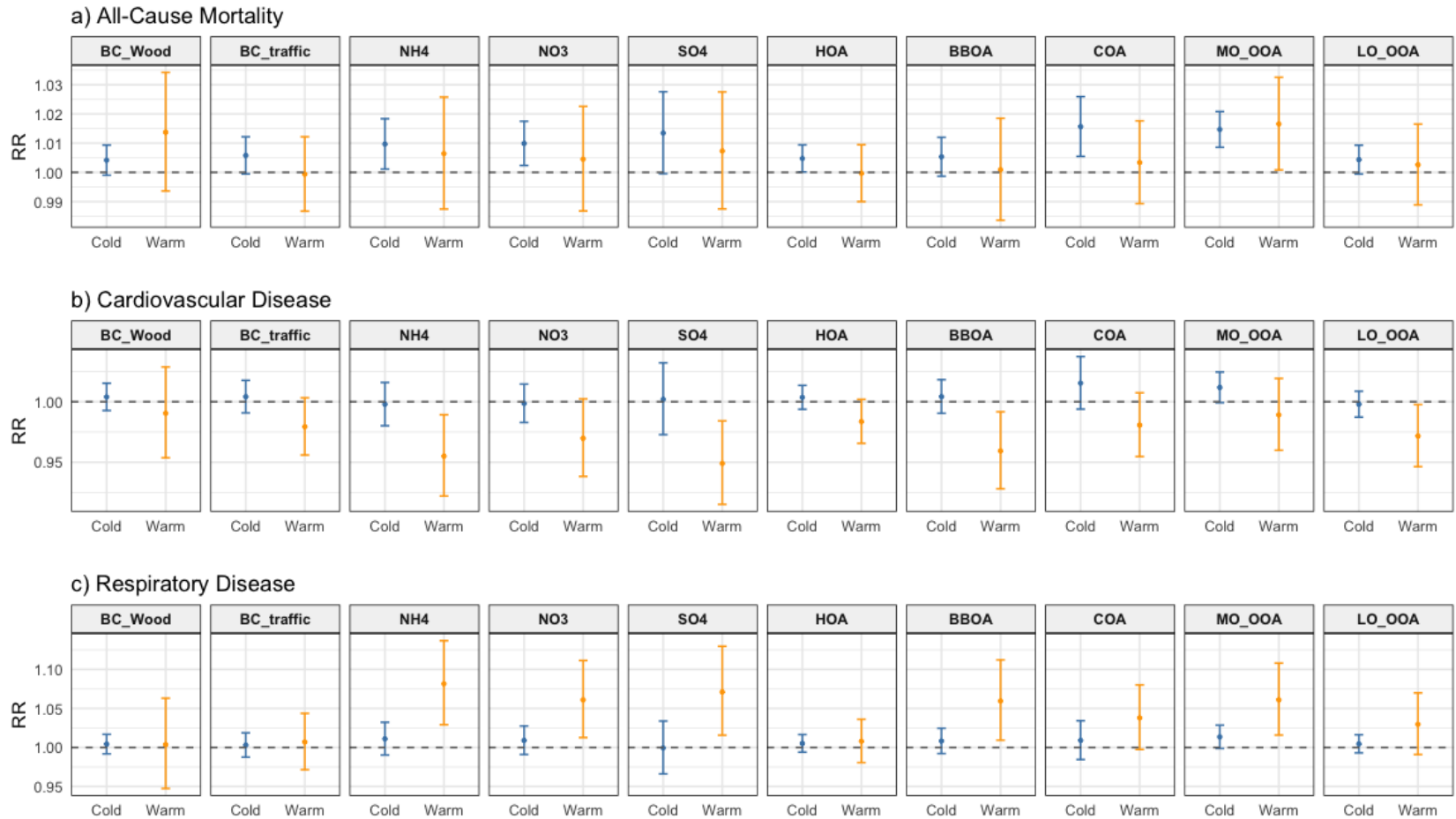

**Figure S4.** Relative risks (RR) and 95% confidence intervals (CI) for the associations between interquartile range (IQR) increases in PM chemical components and daily mortality outcomes (all-cause, cardiovascular, and respiratory), stratified by season (warm: May-October; cold: November-April). Models were adjusted for temperature, relative humidity, long-term and seasonal trends, day of the week, and public holidays.

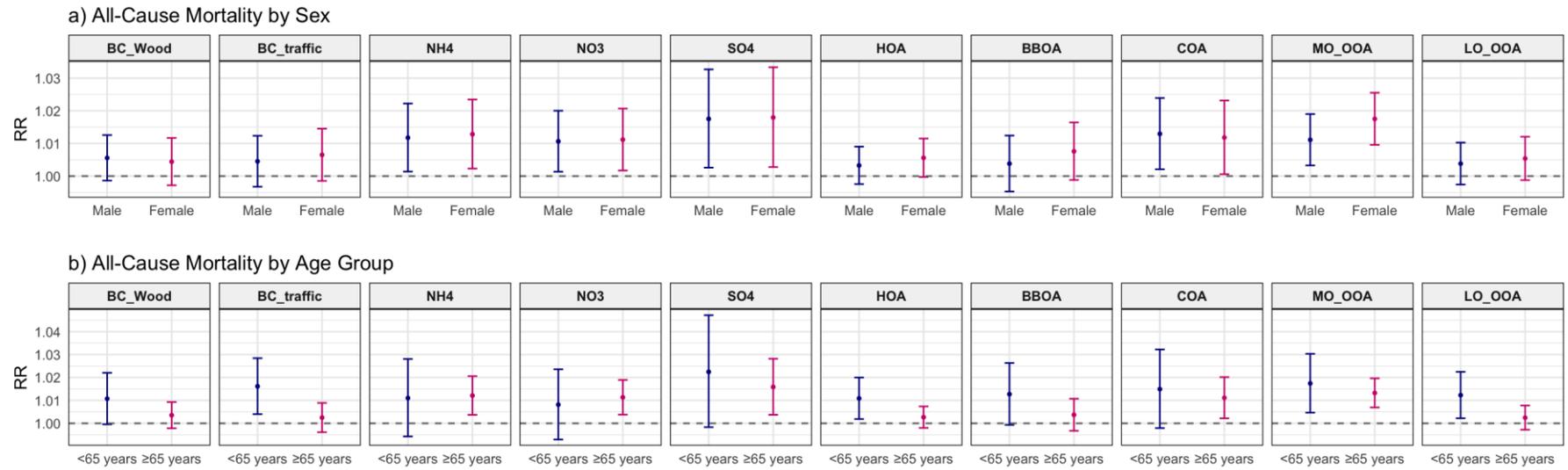

**Figure S5.** Stratified associations between PM components and all-cause mortality by sex (a) and age group (b). Relative risks (RR) and 95% confidence intervals (CI) per IQR increase in PM component concentrations are shown.

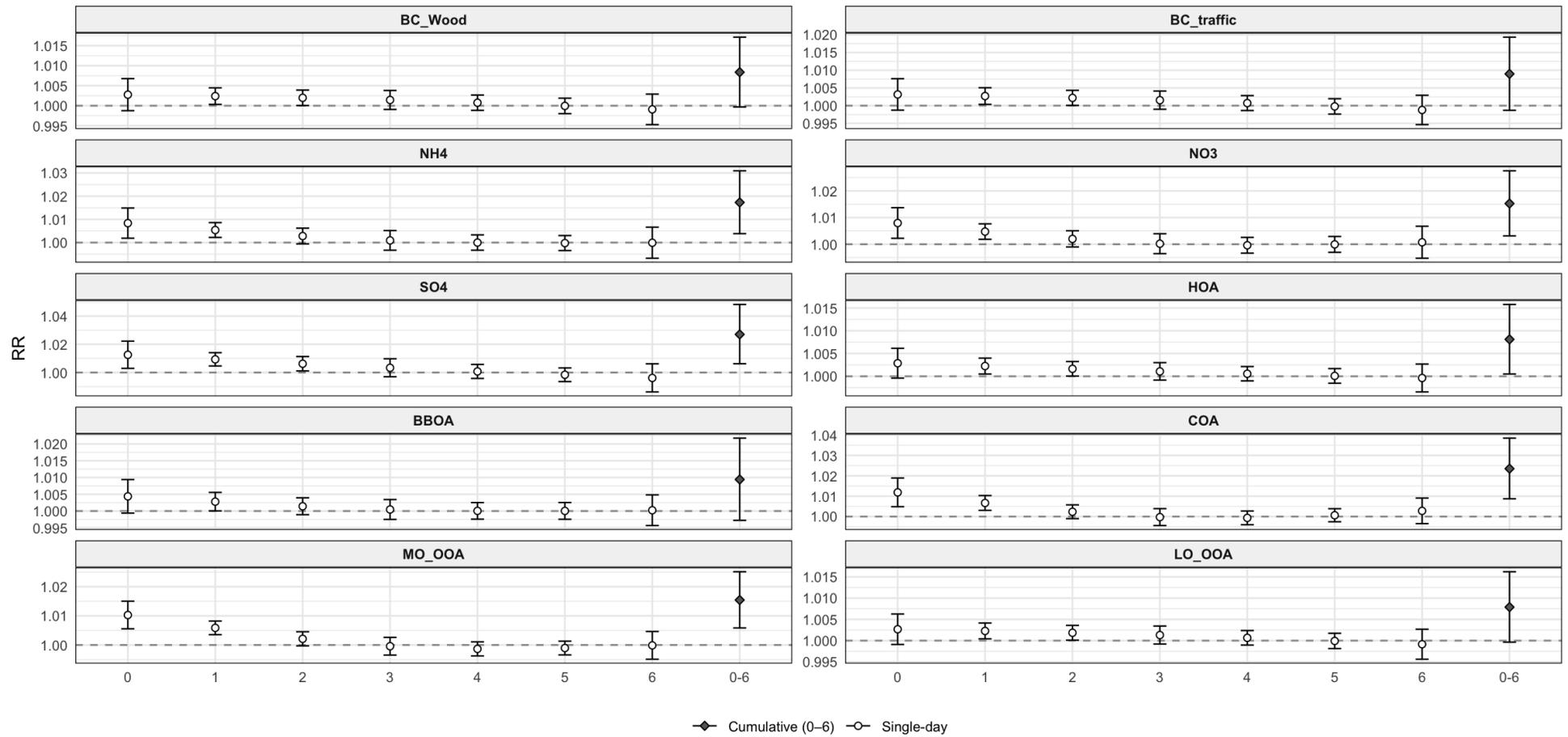

**Figure S6.** Lag-specific (lag 0-6) and cumulative (lag 0-6) relative risks (RRs) with 95% confidence intervals for the associations between PM chemical components and all-cause mortality.

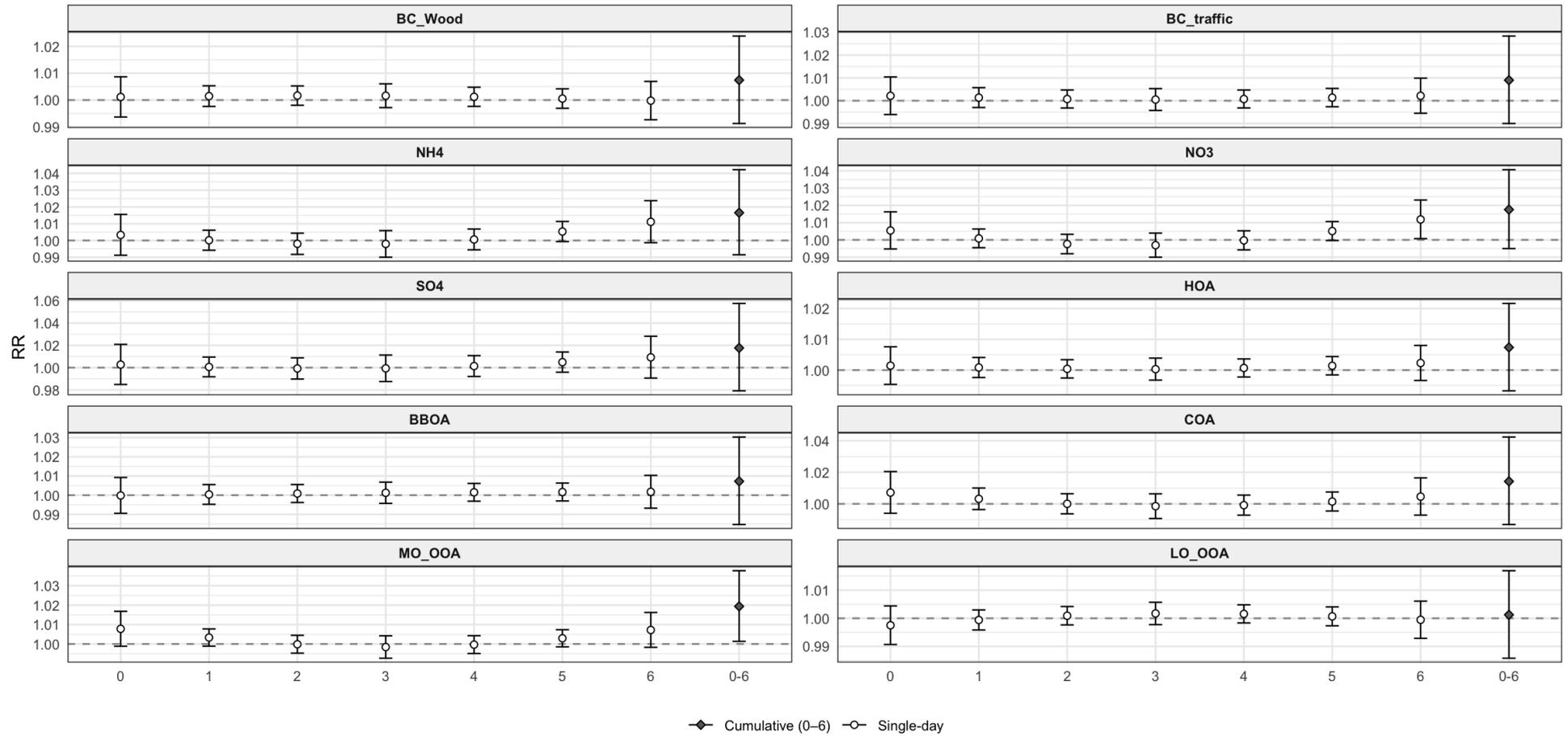

**Figure S7.** Lag-specific (lag 0-6) and cumulative (lag 0-6) relative risks (RRs) with 95% confidence intervals for the associations between PM chemical components and cardiovascular mortality.

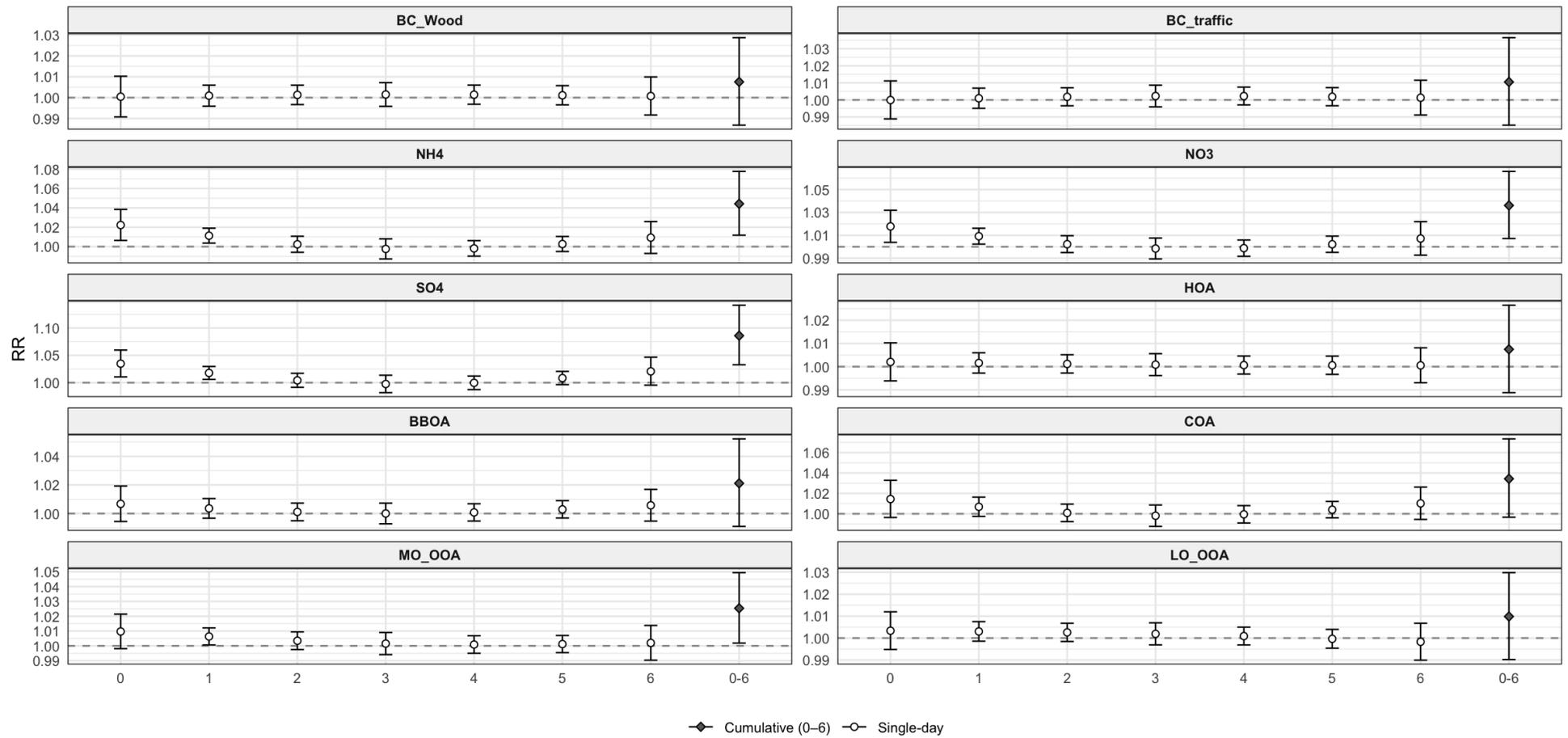

**Figure S8.** Lag-specific (lag 0-6) and cumulative (lag 0-6) relative risks (RRs) with 95% confidence intervals for the associations between PM chemical components and respiratory mortality.

Submitted to Environment International

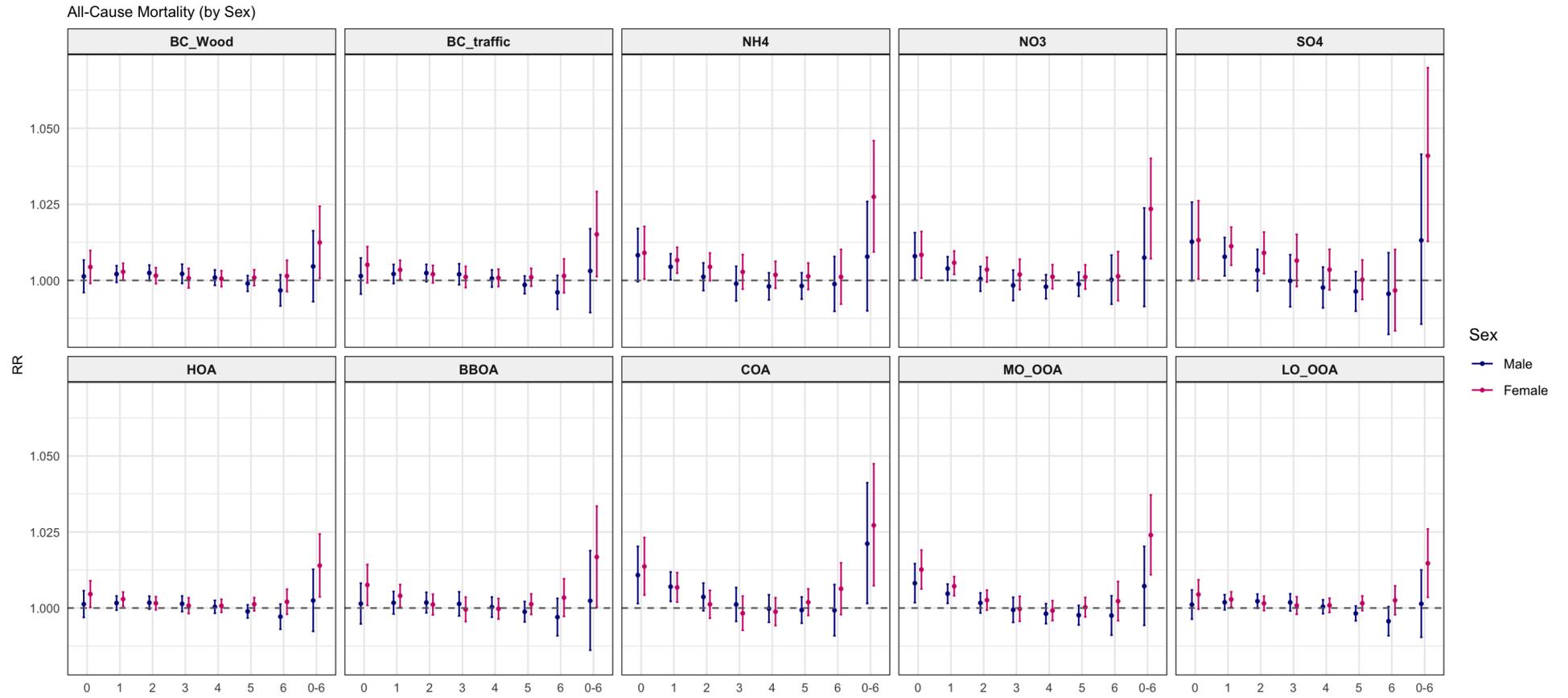

**Figure S9** Lag-specific (0-6) and cumulative (lag 0-6) relative risks (RRs) with 95% confidence intervals for the association between PM components and all-cause mortality, stratified by sex (Male vs. Female).

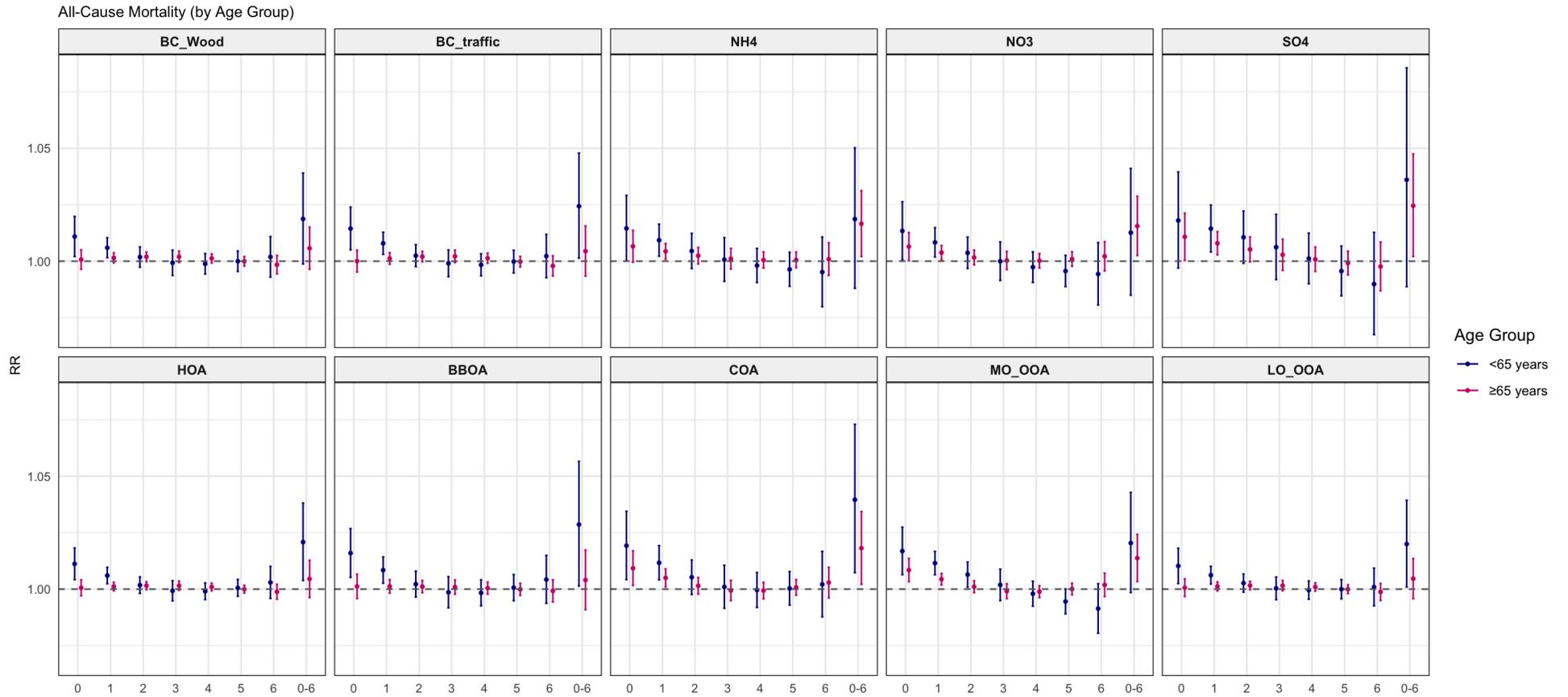

**Figure S10**. Lag-specific (0-6) and cumulative (lag 0-6) relative risks (RRs) with 95% confidence intervals for the association between PM components and all-cause mortality, stratified by age group (<65 years vs. ≥65 years).

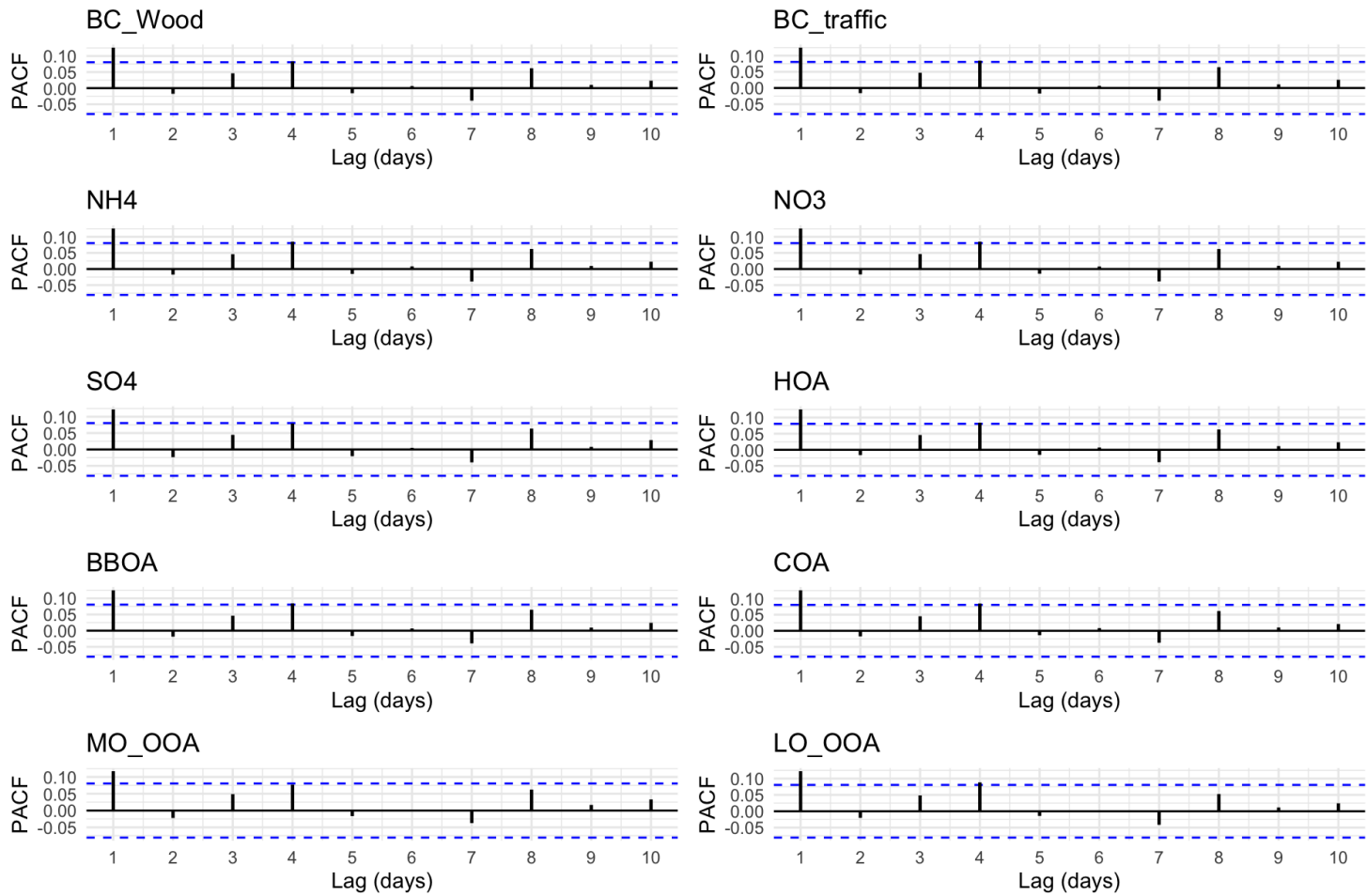

**Figure S11.** Partial autocorrelation functions (PACF) plots of deviance residuals up to lag 10 were examined to assess autocorrelation in each model.

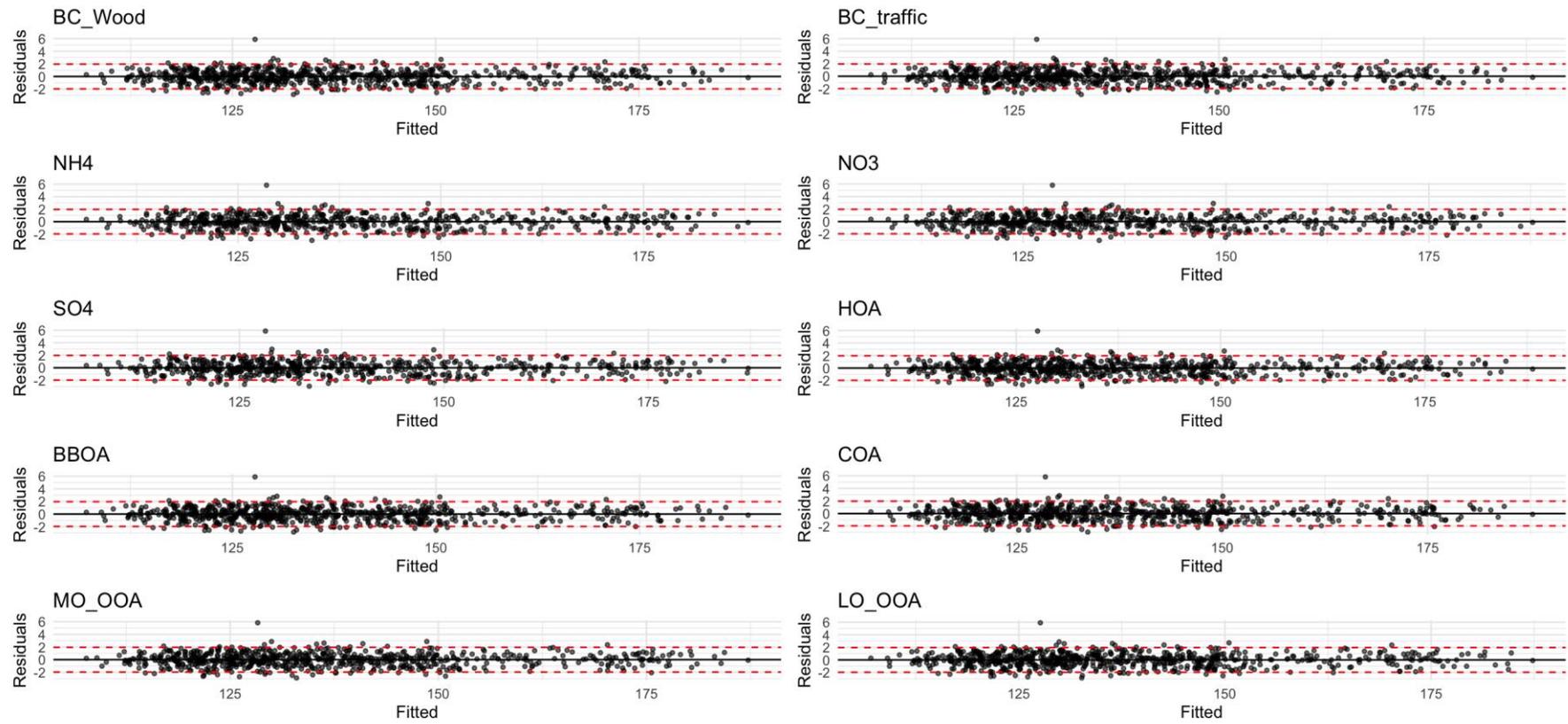

**Figure S12.** Residuals versus fitted values for single-pollutant models to evaluate model fit and variance homogeneity.